\documentclass[fleqn,usenatbib]{mnras}

\usepackage{newtxtext,newtxmath}

\usepackage[T1]{fontenc}

\DeclareRobustCommand{\VAN}[3]{#2}
\let\VANthebibliography\thebibliography
\def\thebibliography{\DeclareRobustCommand{\VAN}[3]{##3}\VANthebibliography}

\usepackage{graphicx}	
\usepackage{amsmath}	

\usepackage{footnote}
\usepackage{comment}
\usepackage{threeparttable}

\usepackage[normalem]{ulem}

\title[ALMA observations of G19.01--0.03 -- {\sc \textit{I}}.]{ALMA observations of the Extended Green Object G19.01--0.03: \\{\sc I}. A Keplerian disc in a massive protostellar system}

\author[G. M. Williams et al.]{G. M. Williams,$^{1,2}$\thanks{E-mail: G.M.Williams@leeds.ac.uk}
C. J. Cyganowski,$^{3}$
C. L. Brogan,$^{4}$
T. R. Hunter,$^{4}$
J. D. Ilee,$^{2}$
P. Nazari,$^{5}$
\newauthor
J. M. D. Kruijssen,$^{6}$
R. J. Smith$^{7}$
and I. A. Bonnell$^{3}$
\\
$^{1}$Centre for Astrophysics Research, Department of Physics, Astronomy and Mathematics, University of Hertfordshire, College Lane, Hatfield, AL10 9AB, UK\\$^{2}$School of Physics \& Astronomy, The University of Leeds, Leeds, LS2 9JT, UK\\$^{3}$Scottish Universities Physics Alliance (SUPA), School of Physics and Astronomy, University of St Andrews, North Haugh, St Andrews, KY16 9SS, UK\\$^{4}$National Radio Astronomy Observatory (NRAO), 520 Edgemont Rd, Charlottesville, VA 22903, USA\\$^{5}$Leiden Observatory, Leiden University, PO Box 9513, 2300 RA Leiden, NL\\$^{6}$Astronomisches Rechen-Institut, Zentrum f\"ur Astronomie der Universit\"at Heidelberg, M\"onchhofstra{\ss}e 12-14, 69120 Heidelberg, DE\\$^{7}$Jodrell Bank Centre for Astrophysics, Department of Physics and Astronomy, University of Manchester, Oxford Road, Manchester, M13 9PL, UK
}

\date{Accepted XXX. Received YYY; in original form ZZZ}

\pubyear{2021}

\begin{document}
\label{firstpage}
\pagerange{\pageref{firstpage}--\pageref{lastpage}}
\maketitle

\begin{abstract}
Using the Atacama Large Millimetre/submillimeter Array (ALMA) and the Karl G. Jansky Very Large Array (VLA), we observed the Extended Green Object (EGO) G19.01--0.03 with sub-arcsecond resolution from 1.05\,mm to 5.01\,cm wavelengths.  
Our $\sim0.4''\sim1600$\,AU angular resolution ALMA observations  
reveal a velocity gradient across the millimetre core MM1, oriented perpendicular to the previously known bipolar molecular outflow, that is consistently traced by 20 lines of 8 molecular species with a range of excitation temperatures, including complex organic molecules (COMs). Kinematic modelling shows the data are well described by models that include a disc in Keplerian rotation and infall, with an enclosed mass of $40-70$\,M$_{\odot}$ (within a 2000\,AU outer radius) for a disc inclination angle of $i=40^{\circ}$, of which $5.4-7.2$\,M$_{\odot}$ is attributed to the disc.  Our new VLA observations show that the 6.7\,GHz Class II methanol masers associated with MM1 form a partial ellipse, consistent with an inclined ring, with a velocity gradient consistent with that of the thermal gas.  The disc-to-star mass ratio suggests the disc is likely to be unstable and may be 
fragmenting into as-yet-undetected low mass stellar companions. Modelling the centimetre--millimetre spectral energy distribution of MM1 shows the ALMA 1.05\,mm continuum emission is dominated by dust, whilst a free-free component, interpreted as a hypercompact H{\sc ii} region, is required to explain the VLA $\sim$5\,cm emission. The high enclosed mass derived for a source with a moderate bolometric luminosity ($\sim$10$^{4}$~L$_{\odot}$) suggests that the MM1 disc may feed an unresolved high-mass binary system.

\end{abstract}




\section{Introduction}
\label{sec:intro}

Both theory and observations suggest that in the low- to intermediate-mass regime (M$_{\ast}<8$\,M$_{\odot}$), protostars accrete material through rotationally supported circumstellar accretion discs and shed excess angular momentum through bipolar outflows \citep[e.g.][]{zinnyorke07}. In the high-mass regime (M$_{\ast}>8$\,M$_{\odot}$), scaling up this process creates an effective pathway for overcoming the hindering effects of high radiation pressure and stellar winds to enable the growth of massive young stellar objects \citep[MYSOs; e.g.][]{krumholz09, kuiper11, klassen16, rosen16,rosen19,meyer18,kuiper18,mignon-risse21}.  MYSOs are indeed commonly observed with active outflows \citep[e.g.][]{beuther02}, however observations of the accompanying circumstellar discs are comparatively lacking. This may be partly attributed to regions of high-mass star formation being several kpc distant and more clustered than their lower mass counterparts. Moreover, the short pre-main sequence lifetimes of high-mass stars \citep[$<1$\,Myr, e.g.][]{mottram11} mean they remain embedded within regions of high extinction in their natal molecular clouds for the duration of their formation  \citep[e.g.][]{kruijssen19,chevance20,kim20}.
High angular resolution studies at (sub)millimetre wavelengths now have the ability to resolve and disentangle thermal emission in these distant, clustered and embedded environments. Most disc candidates around high-mass protostars 
have been observed towards proto-B stars \citep[e.g.][]{cesaroni07,sanchez-monge13,cesaroni14,beltran14,beltran16,girart17,anez-lopez20,jiminez-serra20}, with relatively few candidate 
discs observed towards proto-O stars, e.g.\ AFGL 2591 VLA 3 \citep[][]{jiminez-serra12}, 
NGC 6334 I(N) \citep{hunter14}, IRAS 16547-4247 \citep{zapata15,zapata19}, AFGL 4176 \citep{johnston15,johnston20}, AFGL 2136 \citep{maud18,maud19}, G11.92--0.61 MM1 \citep{ilee16,ilee18}, 
G023.01--0.41 \citep{sanna19}, G345.50+0.35 M, G345.50+0.35 S and G29.96--0.02 \citep{cesaroni17},
with central protostellar masses $\sim$10$-$45\,M$_{\odot}$ and luminosities of $(0.1-5.8)\times10^{5}$\,L$_{\odot}$.

In the search for circumstellar discs around high-mass protostars,
Extended Green Objects \citep[EGOs;][]{cyganowski08, cyganowski09} may represent excellent candidate hosts. EGOs are characterised by extended $4.5\mu$m emission in
\emph{Spitzer} GLIMPSE images \citep{benjamin03,churchwell09} that is thought to trace shocked gas in molecular outflows. 
EGOs are also strongly associated both with radiatively-pumped 6.7\,GHz Class II CH$_{3}$OH masers, which are known to exclusively trace high-mass star formation \citep[e.g.][]{cragg05,billington19,jones20}, and with collisionally-pumped 44 and 25\,GHz Class I CH$_{3}$OH masers \citep[e.g.][]{cyganowski08,cyganowski09,towner17}.  On the whole, EGOs can be inferred to contain one (or more) MYSOs that have active outflows and hence exist in a stage of ongoing accretion.

\begin{figure}
\centering
\includegraphics[scale=.34]{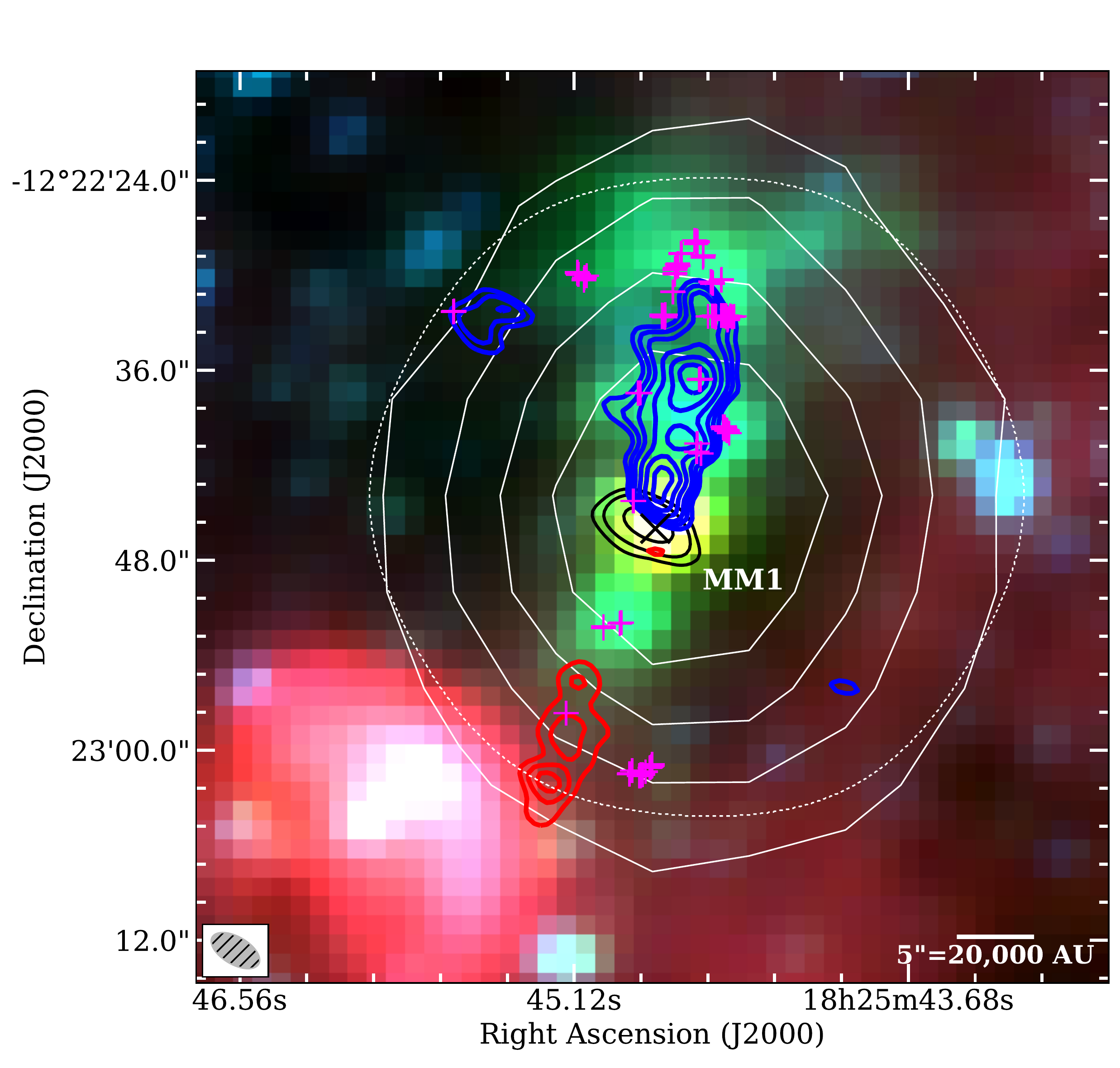}
\caption{\emph{Spitzer} GLIMPSE three-colour image (RGB: 8.0, 4.5, 3.6$\mu$m), overlaid with contours of SMA 1.3\,mm continuum (black: [5, 10, 30]$\times\sigma$, where 1$\sigma=3.5$\,mJy\,beam$^{-1}$) and high-velocity blue- and redshifted $^{12}$CO$(2-1)$ emission (blue: 7.2, 9.6, 12.0, 15.6, 19.2, 22.8 Jy\,beam$^{-1}$\,km\,s$^{-1}$; red: 4.8, 7.2, 9.6 Jy\,beam$^{-1}$\,km\,s$^{-1}$) from \protect\citet{cyganowski11a}.  Positions of 44\,GHz Class I CH$_{3}$OH masers from \protect\citet{cyganowski09} are marked by magenta $+$, and their intensity-weighted 6.7\,GHz Class II CH$_{3}$OH maser position is marked with a black $\times$.  Contours of the ATLASGAL $870\mu$m emission \citep[resolution $18''$;][]{schuller09}
(solid white: [12, 16, 20, 24]$\times\sigma$,
where 1$\sigma = 0.08$\,Jy\,beam$^{-1}$)
and the 30 per cent response level of the ALMA mosaic (dotted white) are also overlaid. The SMA beam is shown at bottom left.}
\label{fig:g19}
\end{figure}

In this series of papers, we present Atacama Large Millimetre/submillimetre Array (ALMA) Cycle 2 observations of the EGO G19.01--0.03 (hereafter G19.01) in Band 7 at 1.05\,mm with $\sim0\farcs4$ resolution,  
the highest resolution observations of this source presented to date.
Figure~\ref{fig:g19} presents an overview of previous (sub)millimetre observations of G19.01.  The millimetre core MM1 appeared
as a single millimetre continuum source in 1.3\,mm Submillimeter Array (SMA) and 3.4\,mm Combined Array for Research in Millimeter-wave Astronomy (CARMA) observations \citep[angular resolution $2\farcs3$ and $5\farcs4$ respectively;][]{cyganowski11a}. 
The 1.3\,mm and 3.4\,mm continuum peaks are coincident with each other, and with 6.7\,GHz Class II CH$_{3}$OH maser emission \citep{cyganowski09,cyganowski11a}. 
Larger scale emission from the surrounding clump is detected 
in both 870\,$\mu$m ATLASGAL \citep[][see Figure~\ref{fig:g19}]{schuller09} and 1.1\,mm Bolocam Galactic Plane Survey \citep[BGPS;][]{rosolowsky10} observations.  From the 1.3\,mm SMA continuum emission, \citet{cyganowski11a} calculated a gas mass of $12-16$\,M$_{\odot}$ for MM1 for $T_{\mathrm{dust}}=130-100$\,K (based on CH$_{3}$CN(J$=$12--11) fitting).  
A remarkably collimated, high velocity bipolar outflow is observed emanating from MM1 in $^{12}$CO(2--1) with the SMA (see Figure~\ref{fig:g19}) and is also detected in HCO$^{+}$(1--0) and SiO(2--1) emission with CARMA.  The lobes of this outflow are also traced by 44\,GHz Class I CH$_{3}$OH masers \citep[Figure~\ref{fig:g19};][]{cyganowski09,cyganowski11a}. 
At the sensitivity level of the SMA and CARMA, MM1 appeared to be lacking in chemical richness.
Complex organic molecules (COMs), defined as carbon-bearing molecules with six or more atoms, are recognised as tracers of high-mass star formation and of hot core MYSOs \citep[e.g.][]{herbst09}.  In the SMA and CARMA observations, only two COMs were detected towards MM1: CH$_{3}$OH and CH$_{3}$CN \citep{cyganowski11a}. 
No 3.6\,cm or 1.3\,cm continuum emission was detected towards MM1 in deep, arcsecond-resolution VLA observations (4$\sigma$ limits 0.12 and 1.04\,mJy\,beam$^{-1}$, respectively), implying a low ionising luminosity (N$_{\rm Lyc}<2.0 \times 10^{45}$\,s$^{-1}$) or very small (diameter $<$80\,AU) \ion{H}{ii} region \citep{cyganowski11b}. The bolometric luminosity of MM1, estimated by fitting the mid-infrared (MIR) to millimetre spectral energy distribution (SED), is moderate, $\sim10^{4}$~L$_{\odot}$ \citep[][see also \citealt{cyganowski11b}]{cyganowski11a}.  In all, these signatures indicate that MM1 is a young millimetre source at an early stage of evolution that is undergoing active accretion, with the highly collimated outflow suggesting that MM1 is an excellent target in the search for discs around MYSOs.

In this paper (Paper {\sc i}), we present an analysis of the kinematics and the centimetre-millimetre wavelength SED of G19.01--0.03 MM1. 
In Section~\ref{sec:observations} we describe the observations and in Section~\ref{sec:results} we present the results for the continuum and the molecular line kinematics.  Section~\ref{sec:discussion} presents our kinematic and SED modelling and discusses the implications of our results for the stability of the MM1 disc and the nature of the central source(s).  Section~\ref{sec:conclusions} summarises our main conclusions.  Throughout, we adopt a near kinematic distance of $4.0\pm0.3$\,kpc, estimated from the NH$_{3}$ LSRK velocity from \citet{cyganowski13} and the Galactic rotation curve parameters from \citet{reid14}.

\section{Observations}
\label{sec:observations}

\subsection{Atacama Large Millimetre/submillimetre Array (ALMA)}
\label{sec:alma_obs}

\begin{table*}
	\small
	\centering
	\caption{Observing parameters of the new data used.}
	\begin{threeparttable}
	\label{tab:properties}
	\begin{tabular}{lccc}
	\hline\hline
	Parameter 					    & ALMA 1.05\,mm  				& VLA 1.21\,cm  				& VLA 5.01\,cm  				\\ \hline
	Observing date 				    & 14 May 2015 				& 11-12 Nov 2013 			& 3 Feb 2014 				\\
	On-source integration time 			    & 44\,min 			& 139\,min 		    & 47\,min			\\
	Number of antennas 			    & 37 						& 25 						& 27						\\
	Antenna configuration 		    & C43-3/(4)					& B 						& BnA 						\\
	Phase Centre (J2000): 			&  		                    &                           &  						\\
	\,\,\,\,\,\,R.A. ($^{\mathrm{h\,m\,s}}$)  & 18:25:44.61$^{a}$               & 18:25:44.80               & 18:25:44.80  \\
	\,\,\,\,\,\,Dec. ($^{\circ}$ $'$ $''$)    & -12:22:44.00$^{a}$              & -12:22:46.00              & -12:22:46.00  \\
	Projected baseline lengths 	    & 20--533\,m 				& 0.14--9.98\,km 			& 0.26--16.54\,km 			\\
		    & 19--508\,$k\lambda$  & 12--825\,$k\lambda$  & 5--330\,$k\lambda$  \\

	Mean frequency$^b$ 		 		    & 285.12\,GHz 				& 24.81\,GHz 				& 5.99\,GHz 				\\
	Mean wavelength$^b$	 		 		    & 1.05\,mm		 			& 1.21\,cm 					& 5.01\,cm 					\\
	Primary beam FWHP$^c$		 	    & n/a (mosaic) 					& 1.8$'$ 				& 7.5$'$ 				\\
	Synthesised beam$^{b}$ 			    & $0\farcs52 \times 0\farcs35$ 	& $0\farcs33 \times 0\farcs22$		& $0\farcs91 \times 0\farcs49$ 	\\
	Beam position angle$^{b,d}$  		    & 88.4$^\circ$ 				& 0.5$^\circ$				& 73.4$^\circ$ 	 			\\
	Maximum Recoverable Scale$^{e}$ & $4\farcs2$ 					& $4\farcs5$ 					& $13\farcs0$ 				\\
	Bandwidth$^{f}$ 					    & $2\times1.875$\,GHz 		& $16\times0.128$\,GHz		& $16\times0.128$\,GHz 						\\
	Channel spacing$^{f}$     		    & 0.977\,MHz 		& 1.000\,MHz	& 1.000\,MHz		\\
    Spectral line RMS noise$^{f,g}$:	& 3.0\,mJy\,beam$^{-1}$ 		& n/a						& n/a						\\

	Continuum RMS noise$^{h}$ 	    & 0.25\,mJy\,beam$^{-1}$ 	& 6.0\,$\mu$Jy\,beam$^{-1}$	& 5.0\,$\mu$Jy\,beam$^{-1}$ \\
	Gain calibrator 			    & J1733--1304 				& J1832--1035				& J1832--1035 				\\
	Bandpass calibrator 		    & J1733--1304 				& J1924--2914				& J1924--2914 				\\
	Flux calibrator 			    & Titan$^{i}$ 					& J1331+3030				& J1331+3030 				\\
	\hline
	\end{tabular}
	\begin{tablenotes}
	    \item[$a$] For the central pointing of the mosaic.
	    \item[$b$] For the continuum image.
	    \item[$c$] At the mean frequency.
	    \item[$d$] Measured East of North i.e. positive in the anti-clockwise direction.
		\item[$e$] Calculated from the fifth percentile shortest baseline (as stated in the ALMA Technical Handbook) and mean frequency, using \texttt{au.estimateMRS} from the analysisUtils Python package.
		\item[$f$] For the two wide spectral windows (ALMA) and continuum spectral windows (VLA); see \S\ref{sec:alma_obs} and \S\ref{sec:vla_obs} for details of narrow spectral windows targeting specific spectral lines.  For the ALMA wideband spws, the Hanning-smoothed spectral resolution is 1.156$\times$ the channel spacing due to online channel averaging in the ALMA correlator.
		\item[$g$] Median value for emission-free channels, imaged with the native channel spacing, for lines presented in this paper. The rms noise is up to $\sim$1.5$\times$ higher in channels with complex emission. 
		\item[$h$] Estimated from emission-free regions within the 30\% response level of the ALMA mosaic. 
		\item[$i$] Using Butler-JPL-Horizons 2012 models.
	\end{tablenotes}
	\end{threeparttable}
\end{table*}

Our ALMA Cycle 2 observations (PI: C.\ Cyganowski; 2013.1.00812.S) mapped G19.01--0.03 at 1.05\,mm with a 7--pointing mosaic.  The ALMA mosaic is $\sim$40\arcsec\/ wide (to the 30 per cent response level), equivalent to $\sim$0.78\,pc at a distance of 4\,kpc.  The coverage of the ALMA mosaic is shown in Figure~\ref{fig:g19}; observing parameters are given in Table~\ref{tab:properties}.  

The ALMA correlator configuration  included seven spectral windows (spws): two broad spws with relatively coarse spectral resolution (Table~\ref{tab:properties}), centred at $\sim$278.2\,GHz and $\sim$292.0\,GHz,
and five narrow spws targeting specific spectral lines. 
Four of the narrow spws have bandwidths of 117.2\,MHz ($\sim$121 km s$^{-1}$) 
and were tuned to cover
H$_{2}$CO $4_{0,4}-3_{0,3}$ at 290.62341\,GHz, DCN (4--3) at 289.64492\,GHz, C$^{33}$S (6--5) at 291.48593\,GHz and $^{34}$SO 6$_{7}$--5$_{6}$ at 290.56224\,GHz; 
the remaining narrow spw has a bandwidth of 468.8\,MHz ($\sim$503 km s$^{-1}$) 
and was tuned to cover N$_{2}$H$^{+}$(3--2) at 279.51176\,GHz. 
For all 5 narrow spws, the (Hanning-smoothed) spectral resolution is 0.244\,MHz.

The data were calibrated using the \textsc{casa} 4.2.2 version of the ALMA calibration pipeline.
Following application of the calibration, the science target fields were split off, and a pseudo-continuum data set constructed from line-free channels.
In our ALMA data, G19.01$-$0.03 MM1 exhibits a line-rich hot core spectrum (\S\ref{sec:lines}); following the approach of \citet{brogan16,cyganowski17}, we selected line-free channels using dirty line$+$continuum cubes.  As in the \citet{cyganowski17} observations with the same tuning, identifying line-free channels in the narrow C$^{33}$S spectral window (spw 3) was problematic due to wide lines and possible absorption features, and we excluded this spw from our aggregate continuum data set.  The total bandwidth used for our final continuum image is $\sim$1.6\,GHz.    

The continuum data were iteratively self-calibrated and the solutions were applied to the line data.
The final continuum image was made using multi-frequency synthesis and Briggs weighting with a robust parameter $R=0$, yielding a synthesised beamsize of 0\farcs52$\times$0\farcs35, equivalent to 2080$\times$1400\,AU at 4\,kpc.
In this paper, we primarily present line data from the two broad spws, as these account for the vast majority of the lines suitable for kinematic analysis (\S\ref{sec:lines}).  
Line image cubes were made with $R=0.5$, and their synthesised beamsizes vary slightly as a function of frequency.  For example, the synthesised beamsize is $0\farcs58\times0\farcs41$ (P.A.$=82.5^{\circ}$) for  g-CH$_{3}$CH$_{2}$OH 16$_{\rm6,10}$--15$_{\rm 6,9}$ (v$_t$=0-0) at $\nu_{\rm rest}=$277.41431\,GHz, and $0\farcs55\times0\farcs40$ (P.A.$=82.5^{\circ}$) for  CH$_{3}$OH (v$_t$=0) 6$_{\rm 1,5}$--5$_{\rm 1,4}$ at $\nu_{\rm rest}=$292.67291\,GHz.  Additional details are given in Table~\ref{tab:properties} for lines in the broad spws.  One line that is suitable for kinematic analysis is serendipitously detected in a narrow spw (that targeting N$_2$H$^+$): CH$_{3}$OH (v$_t$=0) 11$_{\rm 2,10}$--10$_{\rm 3,7}$ \citep[$\nu_{\rm rest}=$279.35189\,GHz, E$_{u}/k_B$=190.9 K;][]{cdms}.
This line was imaged with $R=0.5$ and $\Delta$v = 0.25\,km\,s$^{-1}$; the synthesised beamsize is $0\farcs57\times0\farcs41$ (P.A.$=83.0^{\circ}$) and the rms noise is $\sim$7.2\,mJy\,beam$^{-1}$, measured in emission-free regions of channels with bright emission.
All measurements were made from images corrected for the primary beam response.

\subsection{Karl G. Jansky Very Large Array (VLA)}
\label{sec:vla_obs}

We observed G19.01--0.03 with the Karl G.\ Jansky Very Large Array (VLA) at 1.21\,cm (K band) and 5.01\,cm (C band), under project code 13B-359 (PI: T Hunter).  In this paper, we consider only the continuum and 6.7\,GHz CH$_3$OH maser data.  Observational parameters and continuum image properties are given in Table~\ref{tab:properties}.  Both VLA datasets were calibrated using the \textsc{casa} 4.7.1 version of the VLA calibration pipeline and were Hanning smoothed. 

The K band tuning included sixteen 0.128\,GHz spectral windows for continuum, and narrower spectral windows targeting spectral lines including NH$_{3}$(J=K=$1,2,3,5,6,7$).  
The NH$_{3}$(3,3) line exhibits maser behaviour in other EGOs \citep[e.g.][]{brogan11}, and in our G19.01--0.03 data is strong enough to use for self-calibration.  
The NH$_3$(3,3) line was observed with an 8\,MHz spectral window with 15.625\,kHz channels.
After continuum subtraction in the $u,v$-plane, we performed phase-only self-calibration using the channel with the strongest NH$_{3}$(3,3) emission, and applied the solutions to the continuum data (as well as to the line data, which are not considered further here).
Channels in the continuum spws corresponding to the sky frequencies of targeted spectral lines were flagged prior to continuum imaging to remove line contamination.
The 1.21\,cm aggregate continuum was imaged using multi-frequency synthesis, two Taylor terms to account for the spectral index of the emission across the observed bandwidth and Briggs weighting with a robust parameter $R = 0.5$.  The synthesised beamsize of 0\farcs33$\times$0\farcs22 corresponds to 1320$\times$880\,AU at 4\,kpc.  

The C band tuning included sixteen 0.128\,GHz spectral windows for continuum and one 2.0\,MHz window with 3.906\,kHz ($\sim$0.18 km s$^{-1}$) channels covering the 6.7\,GHz Class II CH$_{3}$OH maser line.  
After continuum subtraction in the $u,v$-plane, the maser data were 
iteratively self-calibrated using the channel with the brightest maser emission; the solutions were also applied to the continuum data.
The maser data were imaged with 0.3\,km\,s$^{-1}$ channels and Briggs weighting with a robust parameter $R = 0$. The resulting image cube has a synthesised beamsize of $0\farcs97\times0\farcs38$ (P.A. 76.3$^{\circ}$) and a 1$\sigma$ rms noise ranging from 1.85\,mJy\,beam$^{-1}$ in line-free channels to 5.35\,mJy\,beam$^{-1}$ in the channel with the strongest emission (see also \S\ref{sec:masers}).
Channels in the continuum spws corresponding to the sky frequencies of emission lines expected in massive star forming regions (including the 6.7\,GHz CH$_3$OH maser line) were flagged prior to continuum imaging to remove line contamination.
The 5.01\,cm aggregate continuum was imaged using multi-frequency synthesis, two Taylor terms to account for the spectral index of the emission across the observed bandwidth, and Briggs weighting with a robust parameter $R=0.5$.  An initial continuum image, made using all data, showed image artefacts from extended emission poorly sampled in our high-resolution data.  To reduce these artefacts and improve our sensitivity to compact emission associated with G19.01--0.03, a $u,v$-range $>20k\lambda$ was applied in making the final continuum image for which properties are quoted in Table~\ref{tab:properties}. All measurements were made from images corrected for the primary beam response.

\begin{table*}
    \centering
    \caption{Observed continuum properties of MM1.}
    \begin{threeparttable}
    \label{tab:leaves}
    \begin{tabular}{lcccccccc}
    \hline\hline
    Tel. / $\lambda$        & Mean   & Source 	     & \multicolumn{2}{c}{J2000.0 Coordinates$^a$}	    & Peak		            & Integ.            &  Source size $^{c}$  		        & Source size 			    \\ 
        & freq.  &	             &  $\alpha$  	& $\delta$ 	                & intensity $^{b}$	    & flux $^{b}$  		& Maj. $\times$ Min. [P.A.]		           &               \\
     	         & (GHz)  &	             &  ($^{\mathrm{h\,m\,s}}$)     & ($^{\circ}$ $'$ $''$)                   & (mJy\,beam$^{-1}$)    & (mJy)             & (\arcsec$\times$\arcsec [$^{\circ}$])                            & (au)                           \\ \hline
    ALMA 1.05\,mm  & 285.115  & MM1       & 18:25:44.782	& -12:22:45.92 				& $266.3$ 				& $303.1$			& $1.15 \times 0.84~[78.7] $ 		        & $4600\times3360$ 		           	\\
    VLA 1.21\,cm    & 24.806   & CM1       & 18:25:44.7821	& -12:22:45.913 				& $0.271\pm0.006$		& $0.295\pm0.011$	& $<0.13 \times <0.03$ 		    & $<520 \times<120$ 		                    	    \\     
    VLA 5.01\,cm    & 5.987 	 & CM1       & 18:25:44.773	& -12:22:46.00 		        & $0.029\pm0.005$		& $0.038\pm0.011$     & $0.4 \times 0.32 ~[75]$	                & $1600\times1280$				 \\ \hline
    \end{tabular}
    \begin{tablenotes}
        \item[$a$] ALMA: peak position, VLA: centroid position from two-dimensional Gaussian fitting, see \S\ref{sec:continuum}.  The number of significant figures reflects a one pixel uncertainty (ALMA) or the statistical uncertainties from the Gaussian fitting (VLA). 
        \item[$b$] ALMA: evaluated within the intensity-weighted second moment size (not the total dendrogram structure). VLA: from two-dimensional Gaussian fits; statistical uncertainties are quoted. 
    	\item[$c$] Major and minor axes sizes, deconvolved from the beam; position angle is measured East of North i.e. positive in the anti-clockwise direction.  ALMA: sizes are calculated from the intensity-weighted second moment (HWHM, using \textsc{astrodendro}) converted to FWHM (and multiplied by $\sqrt{8\ln2}$; shown in Figure~\ref{fig:mom1}). VLA: from two-dimensional Gaussian fitting.  At 1.21\,cm, CM1 is fit as a point source: the reported size is the upper limit from the \textsc{casa imfit} task.  At 5.01\,cm, the fitted size is poorly constrained, with statistical uncertainties of 0\farcs4, 0\farcs09, and 23$^{\circ}$ for the major and minor axes and P.A., respectively.
    \end{tablenotes}
    \end{threeparttable}
\end{table*}

\section{Results}
\label{sec:results}

\subsection{Continuum emission}
\label{sec:continuum}

\begin{figure}
\centering
\includegraphics[scale=0.444]{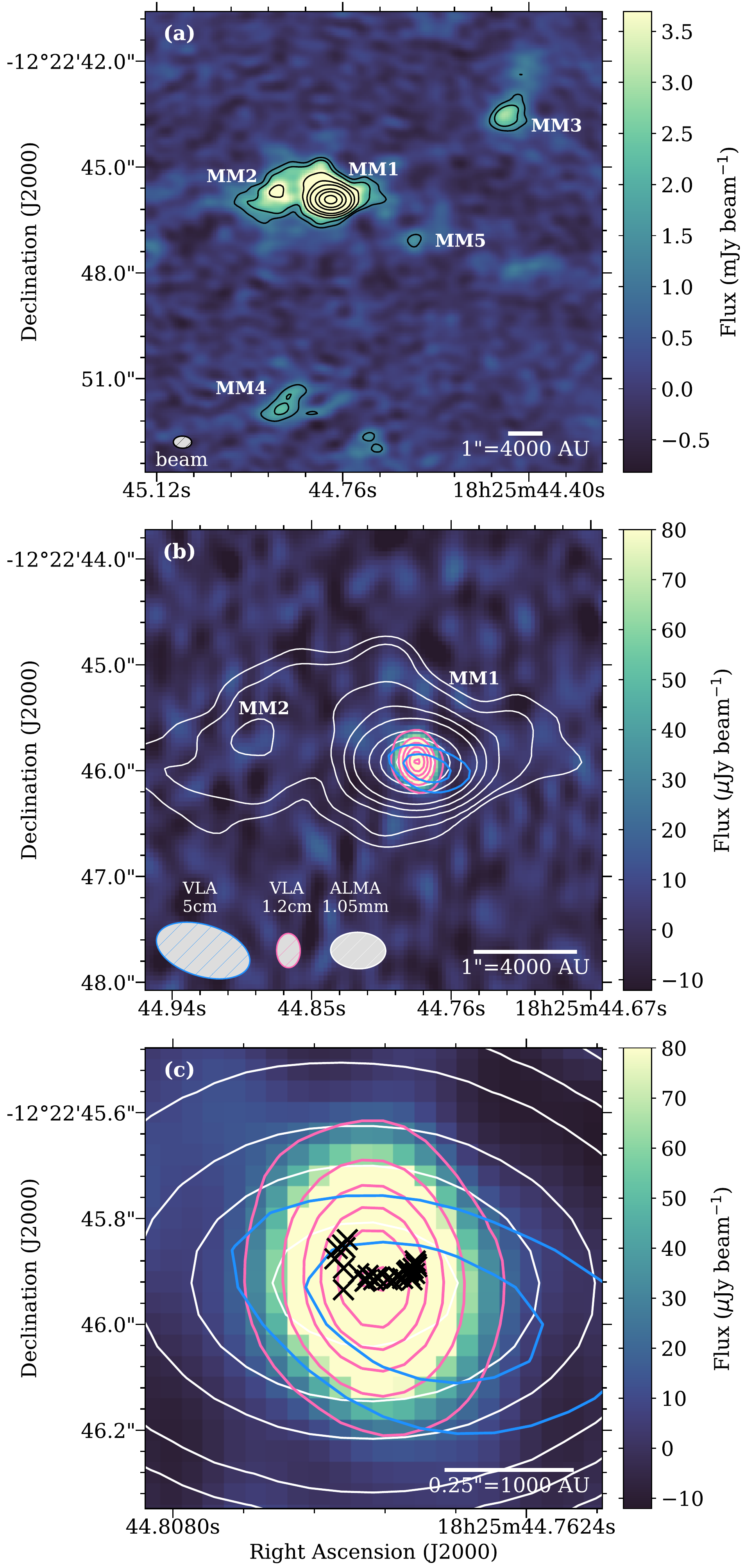}
\caption{(a) ALMA 1.05\,mm continuum image, corrected for the primary beam response, in linear colour scale. To emphasise the low-lying extended emission, the maximum colour scale is limited to 3.7\,mJy\,beam$^{-1}$. The field of view shown is smaller than that of the mosaic, but includes all detected ($\ge$5$\sigma$) emission. Black contours are plotted at 5, 8, 16, 32, 64, 200, 400 and 800$\sigma$, where $\sigma=0.25$\,mJy\,beam$^{-1}$. (b) Zoom view of the VLA 1.21\,cm continuum emission (colour scale, and pink contours in $8\sigma$ steps from 4--44$\sigma$), overlaid with contours of the VLA 5.01\,cm (blue: 4, 5$\sigma$) and ALMA 1.05\,mm (white: 5, 8, 16, 32, 64, 200, 400, 800$\sigma$) continuum emission. The synthesised beams are plotted at lower left, and match the contour colours. MM3, MM4 and MM5 are outside the field of view shown. (c) Zoom view of (b), with new VLA 6.7\,GHz Class II CH$_{3}$OH maser positions (Section~\ref{sec:masers}) plotted as black $\times$.}
\label{fig:leaves}
\end{figure}

The ALMA 1.05\,mm continuum image  shows a strong continuum source at the known position of MM1, sitting within extended emission that is typically $100$ times fainter in surface brightness than the MM1 peak (Figure~\ref{fig:leaves}a,b).  Our deep ALMA continuum image also reveals a few new candidate neighbouring sources, with peak intensities at the level of a few mJy\,beam$^{-1}$. 
Whilst a single 2D Gaussian would reasonably reproduce the MYSO's bright compact emission, it could not account for the non-Gaussian low-lying extended emission. Dendrograms, in contrast, extract structures at regular isocontours of a map, and are typically used in the understanding of hierarchical structure \citep[e.g.][]{goodman09,friesen16,kauffmann17,rigby18,williams18,watkins19,lu20}.  
We use the \textsc{astrodendro} Python package \citep{rosolowsky08dendro} to extract structures across the mosaic area, out to the 30 per cent response level (shown in Figure~\ref{fig:g19}). The algorithm has three free parameters: the minimum isocontour value ($I_{\mathrm{min}}$), the minimum isocontour spacing that separates structures at different isocontours ($\Delta I_{\mathrm{min}}$), and the minimum size of a structure ($n_{\mathrm{pix}}$). We set these parameters to $I_{\mathrm{min}}= 5\sigma_{\mathrm{rms}} = 1.25$\,mJy\,beam$^{-1}$, $\Delta I_{\mathrm{min}}=1\sigma_{\mathrm{rms}} = 0.25$\,mJy\,beam$^{-1}$, and $n_{\mathrm{pix}}=25$\,pixels $\approx n_{\mathrm{pix,beam}}/2$  (where $n_{\mathrm{pix,beam}}$ is the number of pixels in the ALMA synthesised beam) respectively.  
In effect, the wings of the emission profiles of extracted structures are clipped by the algorithm to $I_{\mathrm{min}}$ \citep[see][]{rosolowsky08dendro}. Whilst this does not have a significant impact on strong sources, it can lead to weak point sources being reported with sizes less than a beam. Setting the minimum source size to $\sim$half a beam allows for the extraction of such weak sources despite this effect.
The observed properties of MM1 obtained from the dendrogram analysis are listed in Table~\ref{tab:leaves}. Another four new millimetre sources are extracted in the field (named MM2...MM5, in order of decreasing peak intensity), possibly revealing the early stages of protocluster formation. The properties and nature of these sources will be discussed in Paper {\sc ii}  (Williams et al. in prep.).

Our new 1.21\,cm and 5.01\,cm VLA images are presented in Figure~\ref{fig:leaves}b,c.
As shown in Figure~\ref{fig:leaves}b, the centimetre emission associated with MM1 is isolated and compact, making the use of dendrograms to extract source properties unnecessary. Observed properties of the centimetre emission are estimated from 2D Gaussian fitting, and are presented in Table~\ref{tab:leaves}.
At 1.21\,cm, MM1's centimetre-wavelength counterpart (here called CM1) is strongly detected ($\sim45\sigma$), while at 5.01\,cm only a $\sim5.7\sigma$ detection is made.
The fitted position of CM1 at 1.21\,cm agrees well with the position of the ALMA 1.05\,mm peak (within $<$0\farcs01; Table~\ref{tab:leaves}, Figure~\ref{fig:leaves}b,c).
Interestingly, the 5.01\,cm emission is offset to the southwest by $\sim$0\farcs16 ($\sim$640 AU), $\sim$2.4$\times$ the absolute positional uncertainty of the 5.01\,cm data (estimated as 10 per cent of the geometric mean of the synthesised beam).
Possible interpretations of this offset are discussed in \S\ref{sec:nature_of_MM1}.
No centimetre emission is detected towards the other ALMA millimetre sources in our VLA 1.21\,cm and 5.01\,cm images, to respective $5\sigma$ limits of $30\,\mu$Jy\,beam$^{-1}$ and $25\,\mu$Jy\,beam$^{-1}$.

\subsection{Compact molecular line emission towards MM1}
\label{sec:lines}

Our two broad spectral windows reveal a "forest" of molecular lines towards MM1. In identifying lines, we pay particular attention to those strong and unblended enough for kinematic analysis, and to typical oxygen-bearing COMs expected in hot core sources, which were conspicuously lacking in the previous SMA data (\citealt{cyganowski11a}; Section~\ref{sec:intro}).  We use the JPL \citep{jpl} and CDMS \citep{cdms} catalogues to identify the observed molecular lines. 
As a number of candidate molecular lines may have indistinguishable rest frequencies within the $\sim$1\,km\,s$^{-1}$ spectral resolution of the data, we produce LTE synthetic spectra for each species \citep[e.g.][]{herbst09} using the Weeds extension \citep{maret11} of the CLASS software, accounting for beam dilution.  A line is considered a firm detection if the LTE synthetic spectrum reasonably reproduces the observed brightness temperatures for typical model parameters expected of a hot core, such as line rotational temperatures of $>100$\,K. This approach allowed the identification of some blended lines that were otherwise indistinguishable using rest frequencies alone due to our coarse spectral resolution.  Using this approach, we identify 43 line transitions from 12 different species.  In this paper, we focus on 19 lines from 8 species that appeared strong and unblended enough for kinematic analysis (listed in Table~\ref{tab:lines}); details of other detected lines and a discussion of the chemistry of MM1 will be presented in Paper {\sc ii} (Williams et al. in prep.).

\begin{table*}
	\footnotesize
	\centering
	\caption{Properties of the nineteen spectral lines identified towards MM1 for kinematic analysis in the broad ALMA spectral windows. Arranged by decreasing $E_{u}$/$k_{B}$.}
	\begin{threeparttable}[b]
	\label{tab:lines}
	\begin{tabular}{lcccccc}
	\hline\hline
	Species 						    & Transition  					    & Frequency	        & $E_{u}$/$k_{B}$   & $\sigma^{a}$         & Catalogue$^{b}$ & Kinematics?$^{c}$	   	\\
	                                    &                                   & (GHz)             & (K)               & (mJy\,beam$^{-1}$)         &                   &                    \\ \hline
	CH$_{3}$OH (v$_{t}$ = 0)			& 23$_{4,19}$--22$_{5,18}$		    & 278.96513         & 736.0             & 3.5         & JPL               & Y         \\ 
	CH$_{3}$OH (v$_{t}$ = 0)			& 21$_{-2,20}$--20$_{-3,18}$	  	& 278.48023			& 563.2             & 3.6         & JPL 			    & Y			\\
	CH$_{3}$OH (v$_{t}$ = 0)$^{d}$	    & 17$_{6,12}$--18$_{4,13}$-\,-	    & 291.90814 		& 548.6		        & 4.2        & JPL 			    & Y			\\
	CH$_{3}$OH (v$_{t}$ = 0)			& 18$_{5,13}$--19$_{4,16}$-\,-	    & 278.72314 		& 534.6	            & 3.5       & JPL 			    & Y			\\
	CH$_{3}$OH (v$_{t}$ = 1)			& 10$_{1,10}$--9$_{0,9}$ 		    & 292.51744 		& 418.8             & 4.1       & JPL 			    & Y			\\
	CH$_{3}$OH (v$_{t}$ = 0)			& 14$_{4,10}$--15$_{3,12}$		    & 278.59908			& 339.6	            & 3.5        & JPL			    & Y			\\
	CH$_{3}$CH$_{2}$CN (v = 0)$^{e}$    & 31$_{7,24}$--30$_{7,23}$		    & 278.00758			& 267.8		        & 3.7          & JPL			    & Y			\\
	CH$_{3}$CH$_{2}$CN (v = 0)		    & 31$_{6,25}$--30$_{6,24}$		    & 278.26670			& 253.4		        & 3.7          & JPL			    & Y			\\
	g-CH$_{3}$CH$_{2}$OH $^{f}$		    & 16$_{6,10}$--15$_{6,9}$ (v$_{t}$=0-0)  & 277.41431			& 213.8	    & 3.6	      & JPL			    & Y			\\
	g-CH$_{3}$CH$_{2}$OH		        & 16$_{4,12}$--15$_{4,11}$ (v$_{t}$=0-0) & 278.64299			& 189.7     & 3.6		      & JPL			    & Y			\\
	CH$_{3}$OCH$_{3}$ $^{g}$			& 16$_{1,16}$--15$_{0,15}$ (EA)	    & 292.41225			& 120.3             & 4.1            & JPL		        & Y			\\
	NH$_{2}$CHO						    & 13$_{3,10}$--12$_{3,9}$		    & 277.51403			& 119.8		        & 3.5          & JPL			    & Y			\\
	CH$_{3}$OCH$_{3}$ $^{h}$			& 13$_{2,12}$--12$_{1,11}$ (EE)	    & 291.44307			& 88.0	            & 4.1       & JPL			    & Y	 		\\
	H$_{2}$CO $^{i}$				    & 4$_{2,2}$--3$_{2,1}$			    & 291.94807			& 82.1		        & 4.5          & CDMS			    & Y/N			\\
	CH$_{3}$OCH$_{3}$ $^{j}$			& 12$_{2,11}$--11$_{1,10}$ (EE)	    & 278.40706			& 76.3	            & 3.7       & JPL			    & Y			\\
	CH$_{3}$OH (v$_{t}$ = 0) $^{i}$	    & 6$_{1,5}$--5$_{1,4}$ -\,-		    & 292.67291			& 63.7 		        & 4.1       & JPL			    & Y/N			\\
	$^{13}$CH$_{3}$OH (v$_{t}$ = 0)     & 3$_{2,2}$--4$_{1,3}$ 			    & 291.53662 		& 51.4 		        & 4.1        & CDMS 			    & Y			\\
	$^{13}$CS (v = 0)                   & 6--5                              & 277.45540         & 46.6              & 3.7       & CDMS              & Y/N         \\         
	CH$_{3}$OH (v$_{t}$ = 0)			& 2$_{-2,1}$--3$_{-1,3}$		  	& 278.34222			& 32.9              & 3.7         & JPL			    & Y			\\
	\hline
	\end{tabular}
	\begin{tablenotes}
		\item[$a$] Rms noise measured in emission-free regions of channels with complex emission.
		\item[$b$] CDMS: \url{http://www.astro.uni-koeln.de/cgi-bin/cdmssearch} \citep{cdms}, JPL: \url{http://spec.jpl.nasa.gov/ftp/pub/catalog/catform.html} \citep{jpl}
		\item[$c$] Flag marking lines used for kinematic analysis (Y) or not (Y/N). Those marked with "Y/N" appeared suitable from their isolated spectra, however exhibited extended emission.
		\item[$d$] Blended with CH$_{3}$OH (v$_{t}$ = 0) 17$_{6,11}$--18$_{5,14}$++. Both have the same energy, JPL intensity and frequency.
		\item[$e$] Blended with CH$_{3}$CH$_{2}$CN (v = 0) 31$_{7,25}$--30$_{7,24}$. Both have the same energy, JPL intensity and frequency.
		\item[$f$] Blended with the g-CH$_{3}$CH$_{2}$OH 16$_{6,11}$--15$_{6,10}$ ($v_{t}=0-0$) line, shifted from this reference transition by 0.86\,km\,s$^{-1}$ i.e. $<1$ channel.
		\item[$g$] Blended with three other CH$_{3}$OCH$_{3}$ lines with the same E$_{\mathrm{upper}}$, shifted from this reference transition by 0, 0.17, 0.33\,km\,s$^{-1}$ i.e. $<1$ channel.
		\item[$h$] Blended with three other CH$_{3}$OCH$_{3}$ lines with the same E$_{\mathrm{upper}}$, shifted from this reference transition by 1.55\,MHz $\sim$\,1.6\,km\,s$^{-1}$ i.e. $<2$ channels.
		\item[$i$] Outflow tracing.
		\item[$j$] Blended with three other CH$_{3}$OCH$_{3}$ lines with the same E$_{\mathrm{upper}}$, shifted from this reference transition by 1.7\,MHz $\sim$\,1.8\,km\,s$^{-1}$ i.e. $<2$ channels.
	\end{tablenotes}
	\end{threeparttable}
\end{table*}

\begin{figure*}
\centering
\includegraphics[scale=.35]{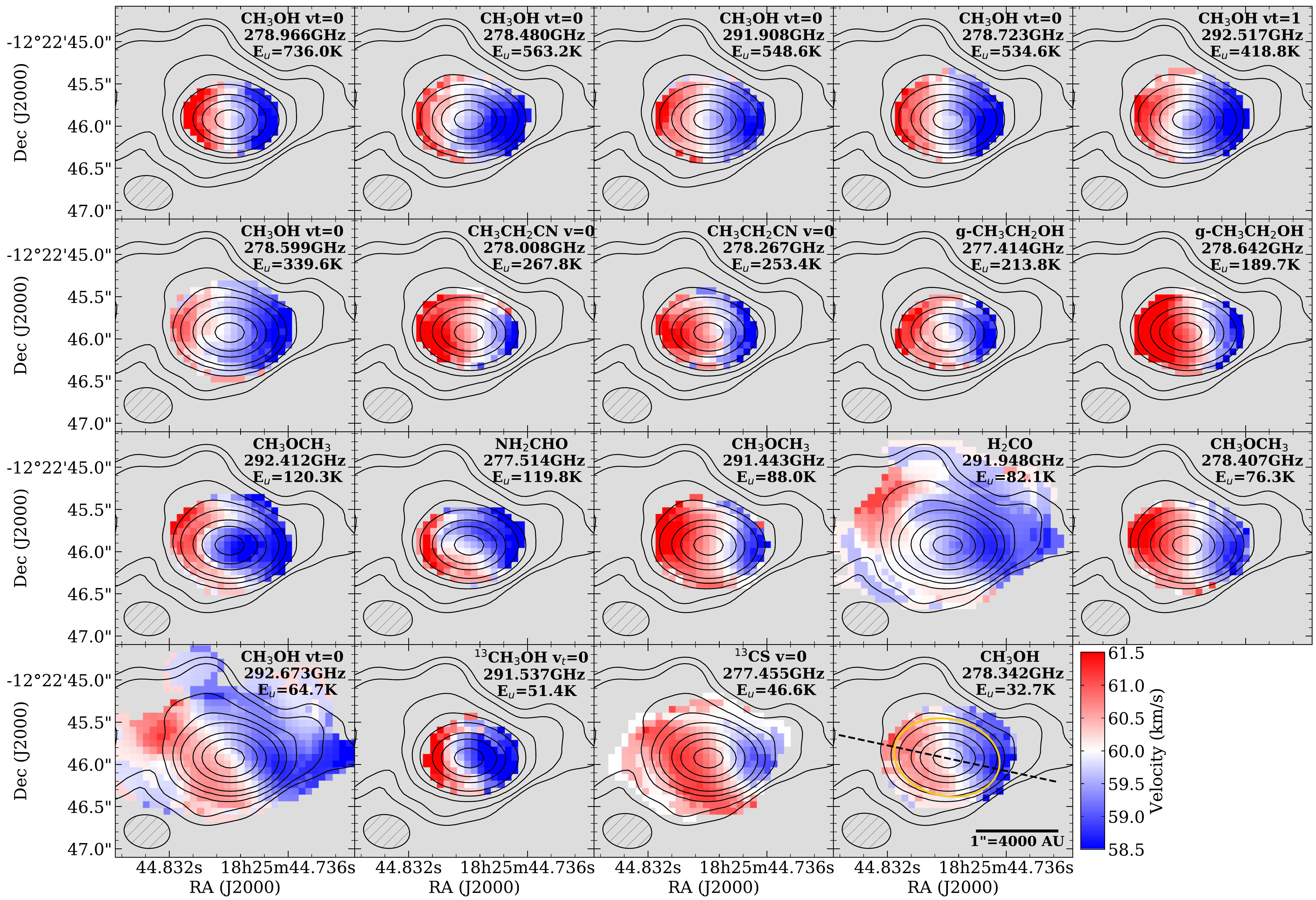}
\caption{Moment one maps of the 19 lines in the wide ALMA spws that appeared unblended and strong enough for kinematic analysis (Table~\ref{tab:lines}), arranged from top-left by decreasing upper level energy in Kelvin.  Each panel is annotated with the species name, rest frequency, and upper level energy.  Masked pixels (line emission $<$5$\sigma_{\mathrm{median}}$, where $\sigma_{\mathrm{median}}=3$\,mJy\,beam$^{-1}$ is the median rms in line-free channels from Table~\ref{tab:properties}) are shown in grey.  ALMA 1.05\,mm continuum contours are shown in black (levels: $5, 8, 16, 32, 64, 200, 400$ and $800\sigma$).  In the bottom right panel, the yellow ellipse shows the size of MM1 measured from the ALMA 1.05\,mm continuum emission (Table~\ref{tab:leaves}), and the dashed black line shows the direction of the PV slices in Figures~\ref{fig:pv} and \ref{fig:pv_ch3oh}.  The synthesised beam is plotted in the bottom left of each panel and a $1''$ scalebar is shown in the bottom right panel.
}         
\label{fig:mom1}
\end{figure*}

Figure~\ref{fig:mom1} and \ref{fig:masers}a present moment one maps of the 19 lines listed in Table~\ref{tab:lines}, and of the narrow-band CH$_{3}$OH (v$_{t}$=0) $11_{2,10} - 10_{3,7}$ line, respectively. 
Most striking is that all species show a consistent velocity gradient across MM1,
with red-shifted emission to the east and blue-shifted emission to the west \citep[with respect to the systemic velocity of $59.9\pm1.1$\,km\,s$^{-1}$;][]{cyganowski11a}.  
The sense of the velocity gradient agrees with that of the 6.7\,GHz Class II CH$_{3}$OH masers imaged by \cite{cyganowski09} (e.g.\ their Figure 5f).
As shown in Figure~\ref{fig:mom1}, the extent of the compact molecular line emission is generally consistent with the measured size of MM1 from the 1.05\,mm continuum emission (shown by the yellow ellipse in the bottom right panel; see also Table~\ref{tab:leaves}).  
As expected, the highest energy lines appear the most compact, with some lower energy lines appearing somewhat more extended.  The three lines marked in the final column of Table~\ref{tab:lines} with ``Y/N'' (H$_{2}$CO($4_{2,3}-3_{2,1}$), CH$_{3}$OH $\mathrm{v}_{t}=0$ ($6_{1,5}-5_{1,4}$) and $^{13}$CS(6--5)) in particular appear in Figure~\ref{fig:mom1} to have more extended morphologies. On larger scales, this H$_{2}$CO and this CH$_{3}$OH line appear to trace MM1's outflow.  To avoid contamination/confusion from larger-scale emission, we exclude these three lines from our kinematic analysis of the candidate disc.

We measure the position angle of the velocity gradient in a similar way to that outlined by \cite{hunter14}, by measuring the slope of the line that connects the most blue-shifted and most red-shifted positions for each line. For the 16 lines in Figure~\ref{fig:mom1} shown to have compact emission (excluding the 3 with more extended emission) and the narrow-band CH$_{3}$OH line in Figure~\ref{fig:masers}a, we find a mean position angle of $78.0 \pm 8.0 ^{\circ}$ (with standard error), in agreement with the position angle of MM1 from the 1.05\,mm continuum of $78.7^{\circ}$. We measure the position angle of the $^{12}$CO(2--1) bipolar outflow observed by \cite{cyganowski11a} with the SMA to be $-18.5\pm5.0^{\circ}$.  
The position angle of the candidate disc is $96.5\pm9.4^{\circ}$ from that of the outflow, consistent with a perpendicular orientation and the observation of a disc-outflow system.

\subsection{6.7 GHz methanol masers}
\label{sec:masers}

\begin{figure}
\centering
\includegraphics[scale=0.489]{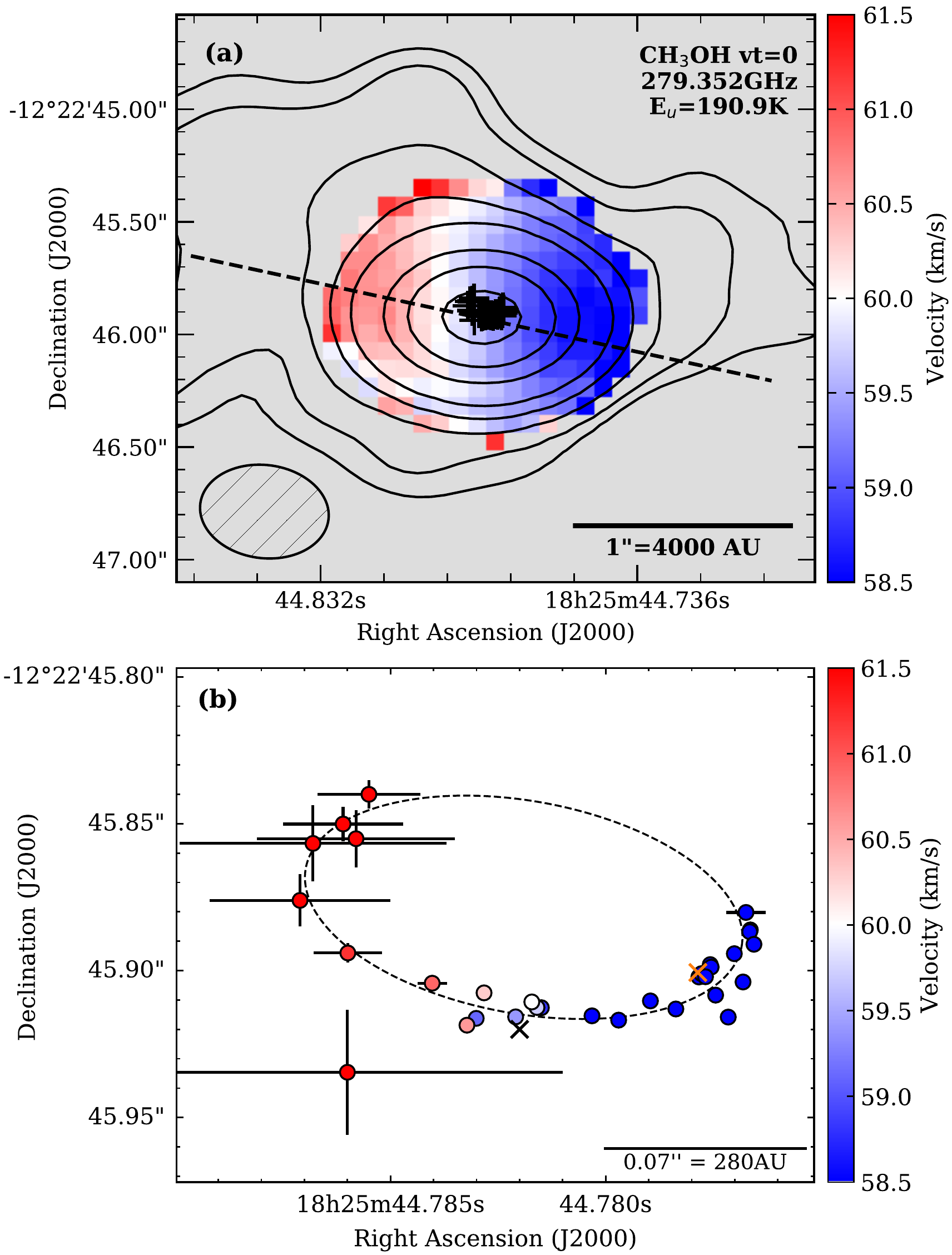}
\caption{(a) Moment one map of the CH$_{3}$OH (v$_{t}$=0) $11_{2,10} - 10_{3,7}$ line ($\nu_{\rm rest}=$279.35189\,GHz, $E_{u}/k_{B}$=190.9\,K) overlaid with ALMA 1.05\,mm continuum contours (black) as in Figure~\ref{fig:mom1}. Masked pixels (line emission $<$5$\sigma$, where $\sigma=7.0$\,mJy\,beam$^{-1}$ is the rms measured in line-free channels) are shown in grey. Black $+$ mark the positions of 6.7\,GHz Class II CH$_{3}$OH masers from Table~\ref{tab:masers}. The dashed line shows the direction of the PV slice in Figure~\ref{fig:pv_ch3oh}.  The synthesised beam is plotted at bottom left, and a $1''$ scalebar is plotted at bottom right. (b) Zoom view showing the fitted positions, with uncertainties, of the masers from Table~\ref{tab:masers}, colour-coded by velocity.  The black and orange $\times$ mark the ALMA 1.05\,mm peak from Table~\ref{tab:leaves} and the intensity-weighted maser position respectively.  The fitted ellipse discussed in \S\ref{sec:masers} is shown as a dashed black line. An $0.07''$ scalebar is plotted at bottom right.}
\label{fig:masers}
\end{figure}

Our new C-band VLA observations provide a higher-resolution view of the 6.7 GHz Class II CH$_3$OH maser emission associated with MM1 \citep[imaged with 2\farcs8 resolution with the VLA by][\S\ref{sec:intro}]{cyganowski09}.
To characterise the morphology and kinematics of the maser emission on scales smaller than the beam, 
we fit the observed emission in each channel with a $>$5$\sigma$ maser detection with a 2D Gaussian using the \textsc{casa imfit} task. 
The rms noise in the maser cube is higher in channels with strong emission due to dynamic range limitations (see also \S\ref{sec:vla_obs}); the $>$5$\sigma$ detection criterion was applied using values of the rms noise measured for each channel within an emission-free region. 
The channels with $>$5$\sigma$ maser emission in our data range in velocity from $53.7-63.0$\,km\,s$^{-1}$ inclusive, and have minimum, maximum and median rms noise values of 1.89, 5.35, and 1.98\,mJy\,beam$^{-1}$. The minimum fitted signal-to-noise is $\sim7.5\sigma$ (corresponding to T$_B=$1051\,K).  
The masers were fit as point sources, i.e.\ with 
the major and minor axes and position angle of the emission fixed to those of the synthesised beam (see Section~\ref{sec:vla_obs}), as we expect the emission to be unresolved \citep[e.g.][]{hunter18,towner21}.  The fitted positions are plotted in Figure~\ref{fig:leaves}c 
and Figure~\ref{fig:masers} 
and tabulated in Table~\ref{tab:masers} along with the fitted peak intensities and statistical uncertainties.

As shown in Figure~\ref{fig:leaves}c, the 6.7\,GHz masers extend over $>$0\farcs1 and coincide with the brightest 1.05\,mm and 1.21\,cm emission.
This can also be seen in Figure~\ref{fig:masers}a and in the zoom view shown in 
Figure~\ref{fig:masers}b, which presents
the fitted maser positions colour-coded by velocity. 
The sense of the velocity gradient seen in the masers is consistent with that of the thermal gas (Figure~\ref{fig:masers}a,b): masers near the $\sim$60\,km\,s$^{-1}$ systemic velocity of MM1 are near the ALMA 1.05\,mm continuum peak, with blue- and red-shifted maser emission to the west and east respectively.
The strongest maser emission is blueshifted (Table~\ref{tab:masers}) and the intensity-weighted position for the group of masers (18$^{\rm h}$25$^{\rm m}$ 44\fs77786 $-$12$^{\circ}$22\arcmin45\farcs90063 (J2000)) is $\sim0\farcs06\sim 240$\,AU from the ALMA 1.05\,mm peak, as shown in Figure~\ref{fig:masers}b. 
We note that our new intensity-weighted maser position is offset from that of \citet{cyganowski09} by $\sim0\farcs15$, within the absolute positional uncertainty of the lower-resolution data estimated as 10 per cent of the geometric mean of the synthesised beam.

\begin{table*}
\centering
\caption{Fitted properties of 6.7\,GHz Class II CH$_{3}$OH maser emission.}
\begin{threeparttable}[b]
\label{tab:masers}
\begin{tabular}{ccccccc}
\hline\hline
\multicolumn{4}{c}{J2000.0 Coordinates}                                      & $I^{b}$                & $dI^{a,b}$              & Velocity      \\ \cline{1-4}
$\alpha$ ($^{\mathrm{h\,m\,s}}$) & $dx^{a}$ ($''$) & $\delta$ ($^{\circ}$ $'$ $''$) & $dy^{a}$ ($''$) & (Jy beam$^{-1}$) & (Jy beam$^{-1}$) & (km s$^{-1}$) \\ \hline
18:25:44.77664        & 0.00165   & -12:22:45.88615            & 0.00046   & 0.6413           & 0.0016           & 53.7          \\
18:25:44.77665        & 0.00083   & -12:22:45.88676            & 0.00023   & 1.5056           & 0.0019           & 54.0          \\
18:25:44.77677        & 0.00690   & -12:22:45.88037            & 0.00221   & 0.3444           & 0.0036           & 54.3          \\
18:25:44.77784        & 0.00021   & -12:22:45.90216            & 0.00006   & 6.5662           & 0.0021           & 54.6          \\
18:25:44.77778        & 0.00017   & -12:22:45.90104            & 0.00005   & 18.657           & 0.0046           & 54.9          \\
18:25:44.77757        & 0.00010   & -12:22:45.89795            & 0.00003   & 21.4148          & 0.0033           & 55.2          \\
18:25:44.77755        & 0.00010   & -12:22:45.89881            & 0.00003   & 17.1632          & 0.0025           & 55.5          \\
18:25:44.77701        & 0.00011   & -12:22:45.89424            & 0.00003   & 13.9754          & 0.0024           & 55.8          \\
18:25:44.77656        & 0.00012   & -12:22:45.89104            & 0.00003   & 14.7786          & 0.0026           & 56.1          \\
18:25:44.77681        & 0.00015   & -12:22:45.90387            & 0.00004   & 11.7347          & 0.0026           & 56.4          \\ \hline
\end{tabular}
\begin{tablenotes}
\item[$a$] Statistical uncertainties from the Gaussian fitting.
\item[$b$] $T_{B}$(K) $\approx 74551 \times I$(Jy~beam$^{-1}$)
\item[] (Only the first ten rows of this table are shown. The full table is available in a machine-readable form in the online journal.)
\end{tablenotes}
\end{threeparttable}
\end{table*}

As illustrated by Figure~\ref{fig:masers}b, our data show that the distribution of the masers forms a partial ellipse, consistent with an inclined ring, centred $\sim$0\farcs04 ($\sim$160 AU) north of the 1.05\,mm continuum peak.
Ring-like 
configurations of Class II CH$_3$OH masers are often observed (most commonly at 6.7\,GHz) and have been attributed to a range of phenomena including 
discs, outflows or outflow cavities and shock interfaces associated with infalling gas \citep[e.g.][]{caswell97,DeBuizer03,bartkiewicz05,bartkiewicz09,bartkiewicz20,torstensson11,sugiyama14,sugiyama16,brogan19}.
An ellipse fitted to the maser positions (excluding the outlier with large uncertainties in Figure~\ref{fig:masers}b) has a size of $0\farcs151\times0\farcs072\sim 600 \times 290$\,AU (major $\times$ minor axis) and a position angle of $\sim80^{\circ}$. 
The position angle of the ring-like maser structure is thus consistent with those of the ALMA 1.05\,mm continuum (Table~\ref{tab:leaves}) and the velocity gradient seen in the thermal molecular lines (\S\ref{sec:lines}), i.e.\ is  
$\sim$perpendicular to the bipolar molecular outflow. 
The physical scale of the maser ring in G19.01 MM1 is similar to those of the smaller rings in the sample of \citet{bartkiewicz09}; for comparison,  
the CH$_3$OH maser ring in the prototypical source G23.657$-$0.127 has a radius of 405\,AU \citep{bartkiewicz20}.
Multi-epoch VLBI imaging has shown that the motions of the G23.657$-$0.127 masers are dominated by radial expansion \citep{bartkiewicz20}, and that a combination of expansion and rotation is required to explain the motions of the 6.7\,GHz CH$_3$OH masers in the EGO G23.01$-$0.41 \citep[which are more widely distributed within an area of $\sim$2800$\times$2800\,AU;][]{sanna10}.
In G19.01 MM1, the consistency in velocity gradient (Figure~\ref{fig:masers}a,b) and position angle between the masers and the thermal line emission suggests a rotational component to the maser motions.
Masers with such motions could plausibly be associated with a (rotating) wide-angle wind at the base of a protostellar jet or the interaction of a wind with the inner regions of the disc \citep[similar to the scenarios proposed for G23.657 and G23.01 by][respectively]{bartkiewicz20,sanna10}, or potentially with an infall/disc interface \citep[as suggested for Cepheus A HW2 by][]{torstensson11}.
Multi-epoch high-resolution observations of the 6.7\,GHz masers in G19.01 MM1 would be required to establish the relative contributions of rotation and expansion or infall to the 3D maser motions to help distinguish among these scenarios.

\section{Discussion}
\label{sec:discussion}

\subsection{Kinematic Modelling of MM1}
\label{sec:pv}

To constrain the enclosed mass, $M_{\mathrm{enc}}$, within the candidate disc,
we produce theoretical position-velocity (PV) diagrams that delineate where emission is expected 
for a thin circumstellar disc in Keplerian rotation and freefall onto the central mass.
Following \cite{cesaroni11}, this is expressed as:
\begin{equation}
    V = \mathrm{V}_{\mathrm{sys}} + \sin{i} \times \left ( \sqrt{GM_{\mathrm{enc}}} \frac{x}{R^{3/2}} + \sqrt{2GM_{\mathrm{enc}}} \frac{z}{R^{3/2}} \right ) \, ,
\label{eq:PV}
\end{equation}
\noindent where $V$ is the velocity along the line of sight, V$_{\mathrm{sys}}$ is the source systemic velocity, $i$ is the line-of-sight inclination angle (where $i=0^{\circ}$ corresponds to a face-on disc), $M_{\mathrm{enc}}$ is the enclosed mass, $x$ and $z$ are respectively co-ordinates along the plane of the disc and the line of sight, and $R=\sqrt{x^2 + z^2}$ is the distance from the centre of the disc and is limited to be between R$_{\mathrm{in}}$ (the inner radius of the disc) and R$_{\mathrm{out}}$ (the outer radius of the disc). The first term in the brackets of equation~\ref{eq:PV} corresponds to the Keplerian disc and the second term to the freefall component, interpreted as infall.

\begin{figure*}
\centering
\includegraphics[scale=.37]{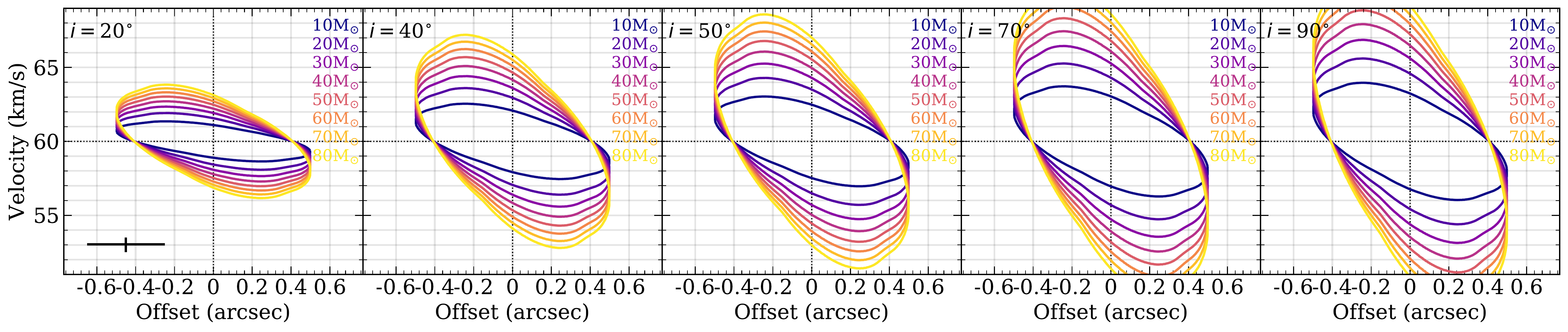}
\caption{Position-velocity (PV) models for a thin Keplerian disc and free-fall motions (equation~\ref{eq:PV}) for M$_{\mathrm{enc}}$ of 10--80\,M$_{\odot}$ and disc inclination angles ($i$) of 20, 40, 50, 70 and 90$^{\circ}$ (from left to right, colours as labelled). In all panels, R$_{\mathrm{out}}$ and R$_{\mathrm{in}}$ are fixed to 2000\,AU and 1700\,AU respectively.  Dotted black lines mark the systemic velocity (60 km s$^{-1}$) and the position of the millimetre continuum peak (zero positional offset).  The angular ($\sim$0.4$''$) and spectral (1.0\,km\,s$^{-1}$) resolution  are shown in the leftmost panel.}
\label{fig:pv_models}
\end{figure*}

\begin{figure*}
\centering
\includegraphics[trim=0 0 78 0,clip,scale=.37]{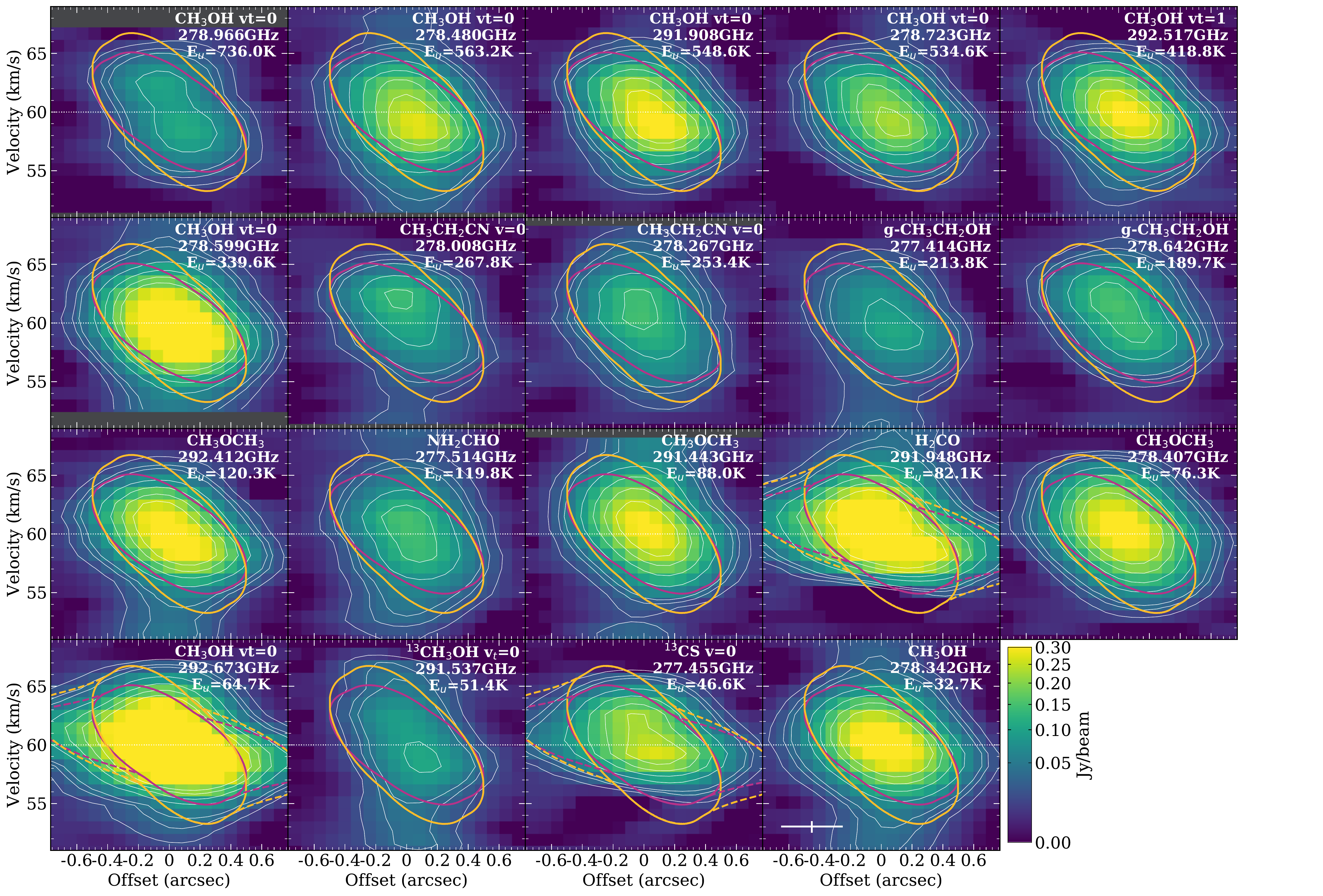}
\caption{Position-velocity (PV) diagrams of MM1 for the 19 lines in the wide ALMA spws selected for kinematic analysis (Table~\ref{tab:lines}), along a slice perpendicular to the direction of the bipolar outflow (see Figure~\ref{fig:mom1}).  The molecule name, rest frequency, and upper energy of the transition are labelled in each panel; panels are arranged from top-left by decreasing upper level energy.  White contours are plotted at 5, 9, 12, 24, 36, 48 and 60$\sigma$ for each line, using the values of $\sigma$ presented in Table~\ref{tab:lines}.  Dark grey pixels (in some panels) indicate regions of PV space outside the imaged sub-cube.  PV models (equation~\ref{eq:PV}) are overplotted as solid lines for $i=40^{\circ}$, R$_{\mathrm{out}}=2000$\,AU, R$_{\mathrm{in}}=1700$\,AU and M$_{\mathrm{enc}}=40$\,M$_{\odot}$ (purple) and M$_{\mathrm{enc}}=70$\,M$_{\odot}$ (orange), representing the range of M$_{\mathrm{enc}}$ that provides a reasonable match to our observed data (\S\ref{sec:pv}).  For the three transitions with extended emission, the coloured dashed lines represent the same model as the corresponding solid line but with a larger R$_{\mathrm{out}} = 3500$\,AU. The angular ($\sim$0.4$''$) and spectral (1.0\,km\,s$^{-1}$) resolution are shown in the bottom right panel. The horizontal dotted white line marks the source systemic velocity of 60 km s$^{-1}$.}
\label{fig:pv}
\end{figure*}

As seen in equation~\ref{eq:PV}, there is a well-known degeneracy between the enclosed mass and disc inclination angle.  For a given R, and pure Keplerian motions or fixed $x$ and $z$, this degeneracy has  the form $V-V_{\rm sys} \sim \sin{i} \sqrt{M_{\mathrm{enc}}}$.  For the case of combined Keplerian rotation and infall, the degeneracy is illustrated by Figure~\ref{fig:pv_models}, where (for example) models with $i=40^{\circ}$ and M$_{\mathrm{enc}}=40\,M_{\odot}$, $i=50^{\circ}$ and M$_{\mathrm{enc}}=30$\,M$_{\odot}$ and $i=70^{\circ}$ and M$_{\mathrm{enc}}=20$\,M$_{\odot}$
all delineate similar regions in PV space.
To constrain $i$, we fit a 2D Gaussian to the moment zero map of each molecular line in Figure~\ref{fig:mom1} with compact emission (excluding the 3 with more extended emission,  \S\ref{sec:lines}, Table~\ref{tab:lines}), and the narrow-band CH$_{3}$OH (v$_{t}$=0) $11_{2,10} - 10_{3,7}$ line in Figure~\ref{fig:masers}a, and calculate $i$ from the deconvolved major and minor axes assuming circular symmetry. 
The inclination angles calculated for these 17 lines range from $i=19-64^{\circ}$, with a median of $i=38^{\circ}$ and a standard deviation of  10$^{\circ}$. If one considered this range to be a reflection of the uncertainty in $i$, then it would correspond to a factor of \begin{math} (\sin ~i)^2 \sim 7.6 \end{math} in the mass estimate.  We caution, however, that the broad range of inclinations estimated from different molecular species may be influenced by molecular abundance, chemistry, optical depth, and/or temperature effects.
The equivalent calculation for $i$ using the deconvolved 1.05\,mm source size from Table~\ref{tab:leaves} yields $i=43^{\circ}$ for the dust emission, in reasonable agreement with the median value of 38$^{\circ}$ from the lines.
While maser-emitting rings are not necessarily co-planar with the thermal gas in the disc \citep[e.g.][]{sugiyama16}, we note that 
the inclination angle of the maser ring, calculated from the fitted size of the ellipse (\S\ref{sec:masers}) assuming circular symmetry, is $i=61^{\circ}$, within the range found for the thermal lines.
Importantly, an intermediate disc inclination angle is also consistent with the observed velocity and morphology of the bipolar molecular outflow driven by MM1 \citep[][see also Figure~\ref{fig:g19}]{cyganowski11a}.

To identify the subset of PV models that provide a reasonable representation of our data, we employ a by-eye approach similar to that of \citet{maud18,maud19} and \citet{ilee18}.
As our primary aim is to constrain the enclosed (and so the stellar) mass,
we fix R$_{\mathrm{out}}$ to 2000\,AU based on the extent of the 1.05\,mm dust continuum emission ($\approx0.5\times$ the geometric mean of the major and minor axes from Table~\ref{tab:leaves}), which is in reasonable agreement with the extent of the compact molecular line emission (Figure~\ref{fig:mom1}).  We fix the inner radius to the beamsize (R$_{\mathrm{in}}=$1700\,AU) to avoid producing features in the model that are not probed by our observations \citep[see e.g.][]{ilee18,jankovic19}. For a given inclination angle $i$, the enclosed mass M$_{\mathrm{enc}}$ is then the only remaining free parameter used to tune the model. We note that since we are limited to a narrow range of radii in the outer disc, the inferred M$_{\mathrm{enc}}$ may be an overestimate since rotation in the outer disc is expected to be super-Keplerian \citep[e.g.][]{kuiper11,kuiper18}.

\begin{figure}
\centering
\includegraphics[scale=.8]{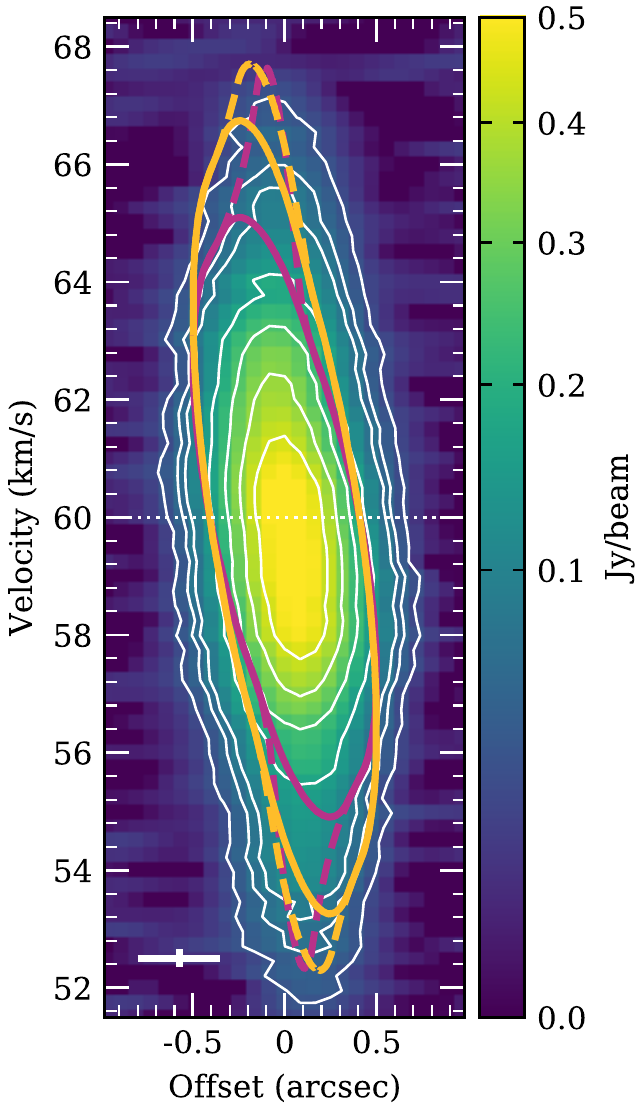}
\caption{Position-velocity (PV) diagram of the CH$_{3}$OH (v$_{t}$=0) $11_{2,10} - 10_{3,7}$ line ($\nu_{\rm rest}=$279.35189\,GHz, $E_{u}/k_{B}$=190.9\,K), imaged with $\Delta$v=0.25 km s$^{-1}$ channels (\S\ref{sec:alma_obs}), along a slice perpendicular to the direction of the bipolar outflow (see Figure~\ref{fig:masers}a). White contours are plotted at 5, 9, 12, 24, 36, 48 and 60\,$\sigma$, where $\sigma = 7.2$\,mJy\,beam$^{-1}$ measured for emission-free regions of line-rich channels.  As in Figure~\ref{fig:pv}, PV models (equation~\ref{eq:PV}) are overplotted as solid lines for $i=40^{\circ}$, R$_{\mathrm{out}}=2000$\,AU, R$_{\mathrm{in}}=1700$\,AU and M$_{\mathrm{enc}}=40$\,M$_{\odot}$ (purple) and M$_{\mathrm{enc}}=70$\,M$_{\odot}$ (orange).  Dashed coloured lines represent models with lower values of R$_{\mathrm{in}}$, of 750\,AU for the 40\,M$_{\odot}$ model (purple) and 1300\,AU for the 70\,M$_{\odot}$ model (orange). The angular ($\sim$0.4$''$) and spectral (0.25\,km\,s$^{-1}$) resolution are shown at bottom left. The horizontal dotted white line marks the source systemic velocity of 60 km s$^{-1}$.}
\label{fig:pv_ch3oh}
\end{figure}

Figures~\ref{fig:pv} and \ref{fig:pv_ch3oh} show observed PV diagrams for the 19 lines in the wide ALMA spws selected for kinematic analysis (Table~\ref{tab:lines}) and the narrow-band CH$_{3}$OH (v$_{t}$=0) $11_{2,10} - 10_{3,7}$ line, respectively, along a slice perpendicular to the direction of the bipolar outflow (see bottom right panel of Figure~\ref{fig:mom1} and Figure~\ref{fig:masers}a). 
The overall shapes of the structures seen in these PV diagrams are generally consistent with each other, with the most notable exceptions in Figure~\ref{fig:pv} being the three lines previously identified as exhibiting more extended emission (low-excitation lines of H$_{2}$CO, CH$_{3}$OH and $^{13}$CS; Table~\ref{tab:lines}, \S\ref{sec:lines}).
As shown in Figure~\ref{fig:pv_ch3oh}, the much higher spectral resolution of the narrowband data ($\sim$4$\times$ better than the wide spws; \S\ref{sec:alma_obs}) probes higher-velocity emission near the central source (i.e.\ at small values of the angular offset). 
Differences between species, and between different transitions of the same species, are seen in the extent and structure of emission in PV diagrams of disc-tracing lines in other MYSOs, due to differences in molecular abundance, excitation, and/or optical depth \citep[e.g.\ NGC 6334I(N)-SMA 1b, G11.92$-$0.61 MM1, and AFGL 4176;][respectively]{hunter14,ilee16,johnston20}. 
Accretion outbursts in the recent past can also affect the appearance/extent of emission of different species \citep[e.g.][and references therein]{wiebe19,jorgensen20}.  
While we see some indications of differences that may be attributable to optical depth (e.g.\ the double-peaked/asymmetric structure and more limited extent of the $^{13}$CH$_3$OH and $E_{u}/k_{B}$=736.0\,K CH$_3$OH lines in Figure~\ref{fig:pv} compared to other CH$_3$OH transitions), their interpretation is limited by the angular resolution of our data.
Disentangling the effects of abundance, excitation, and optical depth would require well-resolved images of a range of molecular species and transitions, and hence higher-resolution (sub)millimetre observations.

Based on the independent evidence for an inclination angle of $\sim$40$^{\circ}$ from the dust continuum and line emission, we focused on the $i=40^{\circ}$ case in our exploration of $M_{\mathrm{enc}}$ parameter space.
For $i=40^{\circ}$, and using the $5\sigma$ intensity contour to guide the eye, we found that models with enclosed masses between 40 and 70\,M$_{\odot}$ (overplotted as coloured lines in Figures~\ref{fig:pv} and \ref{fig:pv_ch3oh}) provide a reasonable representation of the different shapes and extents of the observed emission of the lines along the velocity axis (the extent along the offset axis is fixed to R$_{\mathrm{out}}$, marked by the intersection of the PV models with the horizontal line of the systemic velocity). 
Models with larger outer radii better represent the observed emission of the three lines with more extended emission (low-excitation lines of H$_{2}$CO, CH$_{3}$OH and $^{13}$CS; Table~\ref{tab:lines}), such as R$_{\mathrm{out}}=3500$\,AU (dashed coloured lines in Figure~\ref{fig:pv}). The higher-velocity emission near the central source (i.e. near offset = 0) seen in the CH$_{3}$OH line observed with higher spectral resolution (Figure~\ref{fig:pv_ch3oh}) can be represented by models with smaller inner radii 
(e.g.\ dashed coloured lines in Figure~\ref{fig:pv_ch3oh}) and/or higher enclosed masses, however we cannot distinguish between these parameters with the current data.
We note that while higher angular resolution observations, which better resolve the candidate disc, are required to better measure $i$ directly, we can rule out both very low and very high values of $i$ based on the observed properties of the outflow. 
An edge-on or nearly edge-on disc ($i\sim90^{\circ}$) would imply an outflow in or near the plane of the sky, inconsistent with the very high-velocity $^{12}$CO emission ($>$100 km s$^{-1}$ from the v$_{\rm LSR}$) observed by \citet{cyganowski11a} with the SMA.  At the other extreme, a nearly face-on disc ($i\sim0^{\circ}$) is inconsistent with the extended morphology of both the 4.5\,$\mu$m emission and the high-velocity molecular gas.  
Thus, our kinematic modelling shows that for all plausible inclination angles, a high $M_{\mathrm{enc}}$ ($\ge$15\,M$_{\odot}$ for $i\le80^{\circ}$) is required to describe the observed emission.  In the following sections, we adopt $M_{\mathrm{enc}}=40-70$\,M$_{\odot}$, based on the range of models that best describe our data for the more probable intermediate-inclination ($i=40^{\circ}$) case.

\subsection{SED modelling of MM1}
\label{sec:sed}

\begin{figure}
\centering
\includegraphics[scale=.55]{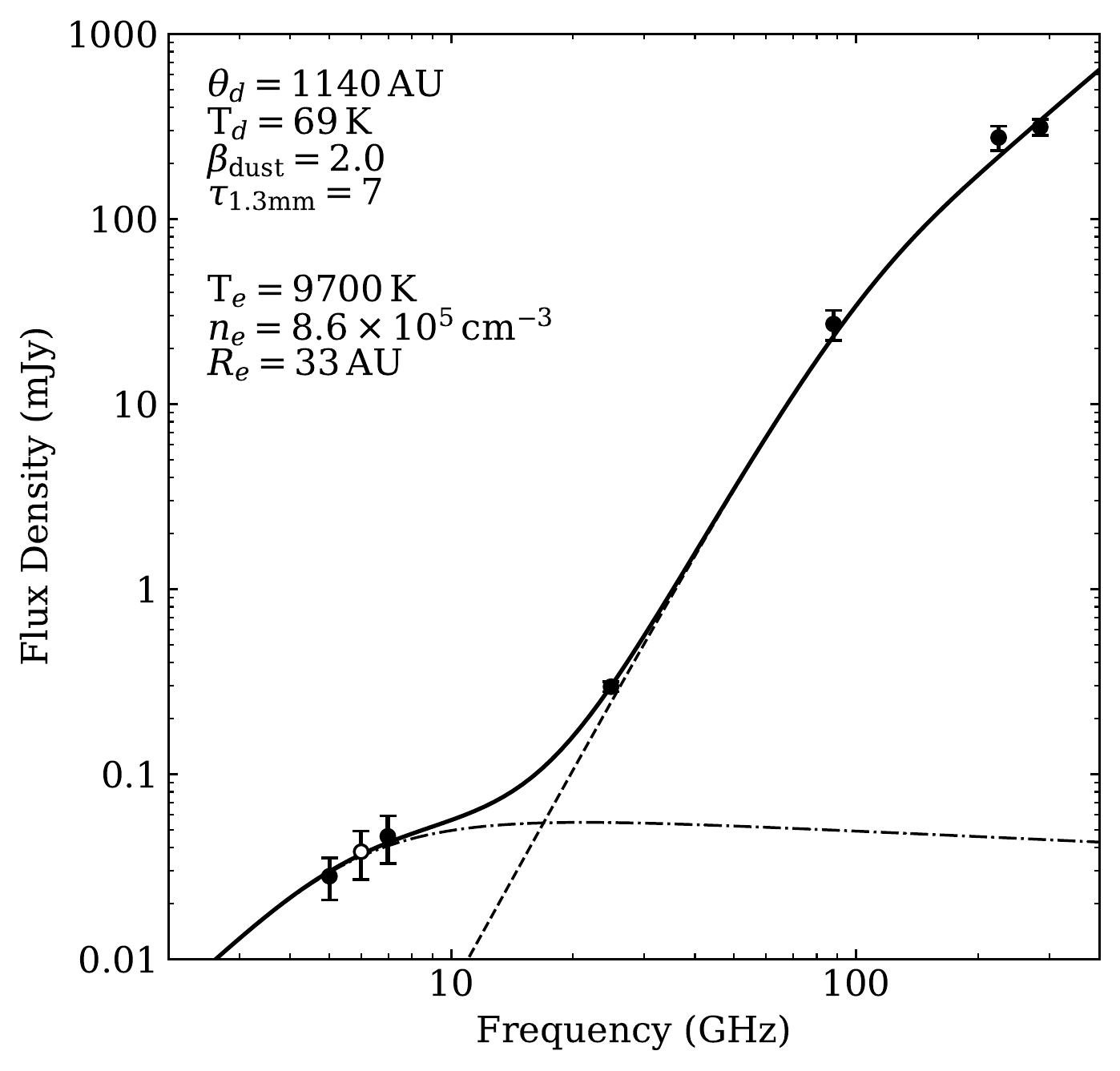}
\caption{Spectral energy distribution of MM1 from cm to mm wavelengths.  Points plotted in black are used for fitting; the 5.01\,cm flux density is overplotted for reference but does not contribute to the fit (Table~\ref{tab:sed_fluxes}, \S\ref{sec:sed}).  Plotted error bars include the statistical uncertainties from Table~\ref{tab:sed_fluxes} added in quadrature with the flux calibration uncertainty (see \S\ref{sec:sed}).   
The best-fitting combined model is overplotted as a solid black line, with the dust component shown as a dashed line and the free-free component as a dot-dashed line.  The parameters of the best-fitting model are printed at upper left.}         
\label{fig:sed}
\end{figure}

In order to constrain the nature and evolutionary state of the central source, we construct and model the centimetre-millimetre wavelength spectral energy distribution (SED) of MM1.  The SED, presented in Figure~\ref{fig:sed}, includes our new VLA 1.21\,cm and 5.01\,cm and ALMA 1.05\,mm data along with the SMA 1.3\,mm and CARMA 3.4\,mm datapoints from \cite{cyganowski11a}.  
To more closely match the $u,v$--coverage of the VLA and ALMA data, we re-imaged the VLA 1.21\,cm data with a $u,v$--range $>20k\lambda$. 
The effect of this adjustment is minimal: the resulting image has a beam size of $0\farcs32\times0\farcs22$ [P.A. $+$0.7$^{\circ}$] and rms noise of $6.0\,\mu$\,Jy\,beam$^{-1}$.
For the purposes of the SED, we measure the integrated flux density at each wavelength by fitting a single 2D Gaussian to the emission, consistent with the approach of \cite{cyganowski11a} for the SMA and CARMA datapoints: the resulting flux densities are presented in Table~\ref{tab:sed_fluxes}.
In Figure~\ref{fig:sed}, error bars represent the statistical uncertainties from the 2D Gaussian fitting, added in quadrature with conservative estimates for the flux calibration uncertainty (5 per cent for the VLA, 10 per cent for ALMA, and 15 per cent for the SMA and CARMA).

As Figure~\ref{fig:sed} illustrates, the millimetre-wavelength emission of MM1 is dominated by dust, but dust emission alone cannot explain the observed shape of the SED. To better constrain the centimetre-wavelength portion of the SED, we imaged the VLA 5.01\,cm data in two halves. These two continuum images have mean wavelengths of 6.00\,cm (5.00\,GHz) and 4.30\,cm (6.97\,GHz), synthesised beams of $1\farcs14\times0\farcs57$ [P.A. 73.8$^{\circ}$] and $0\farcs83\times0\farcs42$ [P.A. 73.4$^{\circ}$], and rms noise levels (1$\sigma$) of $7.8\,\mu$Jy\,beam$^{-1}$ and $6.4\,\mu$Jy\,beam$^{-1}$, respectively. 
The fitted position of CM1 is 18$^{\rm h}$25$^{\rm m}$44\fs758$\pm$0\fs011 $-$12$^{\circ}$22\arcmin45\farcs962$\pm$0\farcs064 (J2000) in the 6.00\,cm image and 18$^{\rm h}$25$^{\rm m}$44\fs7809$\pm$0\fs0060 $-$12$^{\circ}$22\arcmin46\farcs0118$\pm$0\farcs0418 (J2000) in the 4.30\,cm image (see discussion in \S\ref{sec:nature_of_MM1}).
The flux densities measured from these images are presented in Table~\ref{tab:sed_fluxes} and Figure~\ref{fig:sed}:
the signal to noise of CM1 is $\sim3.7\sigma$ in the 6.00\,cm image and $\sim5.8\sigma$ at 4.30\,cm. 
The centimetre-wavelength spectral index calculated from these datapoints, $\alpha_{\rm 6.00cm-4.30cm}=1.5\pm1.1$ (S$_{\nu}$ $\propto$ $\nu^{\alpha}$), is consistent with moderately optically thick free-free emission from a hypercompact (HC) \ion{H}{ii} region or ionised jet \citep[e.g.][]{moscadelli16, purser16,yang19}.

\begin{table}
\centering
\caption{Integrated continuum flux densities evaluated from 2D Gaussian fitting, and used in the SED fit.}
\label{tab:sed_fluxes}
\begin{threeparttable}[b]
\begin{tabular}{cccc}
\hline
Wavelength        & Frequency & Flux density$^a$    & Imaged $u$,$v$--coverage$^b$   \\
                  & (GHz)     & (mJy)           & (k$\lambda$)          \\ \hline
6.00\,cm$^c$      & 5.00      & $0.028\pm0.007$ & $20-276$ \\
5.01\,cm$^d$    & 5.99      & $0.038\pm0.011$ & $20-330$ \\
4.30\,cm          & 6.97      & $0.046\pm0.013$ & $20-385$ \\
1.21\,cm          & 24.81     & $0.296\pm0.011$ & $20-825$ \\
3.4\,mm$^{e}$     & 88.0     & $27\pm3$        & $1.5-36.5$       \\ 
1.3\,mm$^{e}$     & 225.1    & $275\pm7$       & $7-88$      \\
1.05\,mm          & 285.12    & $313.1\pm0.5$   & $19-508$   \\\hline
\end{tabular}
\begin{tablenotes}
\item[$a$] Uncertainties are the statistical uncertainties from the Gaussian fitting.
\item[$b$] Projected baseline ranges used in the images used to construct the SED, including the adjustments described in \S\ref{sec:vla_obs} and \S\ref{sec:sed}.   
\item[$c$] Source size fixed to the synthesised beamsize for fitting due to low S/N ($\sim$3.7).
\item[$d$] From Table~\ref{tab:leaves}: this point is plotted in Figure~\ref{fig:sed} but not included in the SED fit, due to the inclusion of the 4.30\,cm and 6.00\,cm flux densities (see \S\ref{sec:sed}). 
\item[$e$] From \cite{cyganowski11a}
\end{tablenotes}
\end{threeparttable}
\end{table}

We thus model MM1's centimetre-millimetre wavelength SED using a two component model \citep[similar to the approach of][]{hunter14,brogan16,ilee16}, where the dust emission is described by a single-temperature modified greybody function \citep[e.g.][]{gordon95,rathborne10}, and the free-free emission by a Bremsstrahlung model \citep{olnon75} interpreted as an HC \ion{H}{ii} region. Since the centimetre source is marginally resolved at best, the only geometry we model is that of a uniform sphere. This combined model has seven free parameters: the electron density ($n_{e}$), radius (R$_{e}$), and electron temperature (T$_{e}$) of the HC \ion{H}{ii} region, and the angular diameter ($\theta_{d}$), temperature (T$_{d}$), grain emissivity index ($\beta$), and reference opacity ($\tau_{\mathrm{1.3mm}}$) of the dust emission. 
Since we have only six independent datapoints, we employ a two-stage process to explore the parameter space of the combined model: 1) for each point in a grid of R$_{e}$ and n$_{e}$, we fit for the five remaining free parameters; 2) adopting the best-fit \ion{H}{ii} region parameters from step 1 (based on the reduced $\chi^2$), we fit for the dust opacity and angular diameter for each point in a grid of $\beta$ and dust temperature. 
The best-fitting model, shown in Figure~\ref{fig:sed}, has an \ion{H}{ii} region radius R$_e$=33\,AU \citep[consistent with the radius upper limits of 36\,AU and 40\,AU from][]{cyganowski11b}, an electron temperature of 9700\,K, and an electron density of 8.6$\times$10$^{5}$\,cm$^{-3}$; the dust component is optically thick at millimetre wavelengths ($\tau_{\mathrm{1.3mm}}$=7), with a diameter of 1140\,AU, $\beta$=2.0, and T$_d$=69\,K. 
We note that this model is not unique, as there are degeneracies between model parameters that cannot be broken without additional high-resolution data: in particular, the dust temperature T$_d$ 
is effectively unconstrained.
By including the SMA 1.3\,mm and CARMA 3.4\,mm data, we are, however, able to obtain a good constraint on $\beta$, which is important for constraining the relative contributions of the dust and free-free components at intermediate frequencies.  
In sum, our modelling demonstrates that emission from both dust and ionised gas is required to explain the observed SED of MM1.  Our results support a picture in which the ionised gas is confined to a very small HC \ion{H}{ii} region, suggesting that the \ion{H}{ii} region may be gravitationally "trapped" by an accretion flow \citep[e.g.][]{keto03,keto07}, consistent with other evidence for ongoing accretion (\S\ref{sec:intro}).

\subsection{Disc properties and stability}
\label{sec:disc_prop}

Our SED modelling confirms that the ALMA 1.05\,mm emission of MM1 is dominated by thermal dust emission, with dust accounting for 99.99\% of the 1.05\,mm flux density in our best-fitting model.
We estimate the mass, M$_{\mathrm{gas}}$, of MM1 from its 1.05\,mm integrated flux density in Table~\ref{tab:leaves}, assuming isothermal dust emission and correcting for the dust optical depth  \citep[e.g.][]{cyganowski11a,cyganowski17}:
\begin{equation}
	\mathrm{M_{\mathrm{gas}}} = \frac{d^{2}\,R\,S_{\nu}\,C_{\tau_{\mathrm{dust}}}}{\kappa_{\nu}\,B_{\nu}(T_{\mathrm{dust}})} \, ,
	\label{eq:mdust}
\end{equation}
\noindent where $d$ is the distance, $R$ is the gas-to-dust mass ratio (assumed to be 100), $S_{\nu}$ is the  1.05\,mm integrated flux density, $\kappa_{\nu}$ is the dust opacity at 1.05\,mm, $B_{\nu}(T_{\mathrm{dust}})$ is the Planck function, $T_{\mathrm{dust}}$ is the dust temperature, and $C_{\tau_{\mathrm{dust}}}$ is a correction for the dust optical depth:
\begin{equation}
    C_{\tau_{\mathrm{dust}}}= \tau_{\mathrm{dust}}/(1-e^{-\tau_{\mathrm{dust}}}) \, ,
\end{equation}
\noindent where $\tau_{\mathrm{dust}}$ is estimated as:
\begin{equation}
    \tau_{\mathrm{dust}} = -\ln\left(1-\frac{T_{b}}{T_{\mathrm{dust}}}\right) \, .
\end{equation}
\noindent As in \citet{cyganowski17}, we assume $\kappa_{\rm 1.05mm}=$1.45\,cm$^2$\,g$^{-1}$ \citep[for grains with ice mantles in high gas density environments;][]{ossenkopf94}.
We estimate $T_{b}$ as the mean Planck brightness temperature across the intensity-weighted second moment size of MM1 (10.9\,K).  On this sizescale, for temperatures $T_{\mathrm{dust}}=100-130$ K \citep[based on the CH$_3$CN fitting in][]{cyganowski11a}, we calculate dust optical depths $\tau_{\mathrm {dust}}$\,$\sim$\,0.09$-$0.12 and a gas mass M$_{\mathrm{gas}}$\,$\sim$\,5.4$-$7.2\,M$_{\odot}$ for MM1.  We note, however, that these estimates do not capture expected variations in temperature and opacity within a disc \citep[e.g.][]{carrasco19,jankovic19}, and that our adopted temperature range may underestimate the temperature on the smaller sizescales probed by our ALMA observations.

With these caveats in mind, we can consider the stability of the star-disc system in MM1 based on the physical parameters of the star and circumstellar disc derived from our observations.
The disc-to-star mass ratio ($M_{d}/M_{\ast}$) is often used as a proxy for the gravitational (in)stability of a disc, with unstable discs typically having values $>0.1$ \citep[e.g.][and references therein]{kratter16}. 
Such a stability criterion relies on the evaluation of a reliable disc mass; however, simple mass estimates based on millimetre continuum emission (as in equation~\ref{eq:mdust}, or assuming optically thin emission) can underestimate the disc mass, as they do not account for variations in temperature, dust opacity, and dust optical depth within the disc \citep[e.g.][]{johnston15,forgan16}. 
\cite{forgan16} computed semi-analytical disc models for five massive young stars with candidate discs from the literature, allowing a direct comparison of the "true" disc mass in each model with the observed mass that would be inferred from millimetre observations (adopting common assumptions, including optically thin emission). 
For this sample, which included star-disc systems broadly similar in properties to MM1, \citet{forgan16} found that a system with a "true" disc-to-star mass ratio of $\sim0.2$, indicative of instability, could have an \emph{observed} disc-to-star mass ratio of $\leq0.05$. 

For MM1, attributing the $M_{\mathrm{gas}}$ estimated above to the disc and estimating the stellar mass as $M_{\ast}=M_{\mathrm{enc}}-M_{d}\simeq33-65$\,M$_{\odot}$, we find $M_{d}/M_{\ast}\sim0.08-0.22$. 
At the cusp of stability (i.e. $M_{d}/M_{\ast}=0.1$), and for $M_{d}=5.4$\,M$_{\odot}$ (the low end of our range) to check for the most stable case, an enclosed mass of $\gtrsim$59\,M$_{\odot}$ is required for the system to be stable.  If the true disc mass is in fact higher than our observational estimate, as discussed above, this would have the effect of making the disc more unstable, in turn requiring an even larger enclosed mass for stability. 
We note that if the dust temperature on the sizescales probed by our ALMA observations is higher than our adopted temperature range, this would have the effect of decreasing our estimate of $M_{d}$ and pushing the system towards stability.
We calculate the minimum $T_{\mathrm{dust}}$ for which the system would be stable (for isothermal dust emission, and other assumptions as outlined above): 112\,K for $M_{\rm enc}=70$\,M$_{\odot}$ and 186\,K for $M_{\rm enc}=40$\,M$_{\odot}$.
Subsequent higher-resolution studies of the other EGO studied by \citet{cyganowski11a}, G11.92$-$0.61, suggest that a resolution-dependent temperature increase of $\sim$ 60-70 K is possible: \citet{ilee16} found that two components with temperatures of $\sim$150 and 230\,K were required to model the CH$_3$CN spectra of the G11.92$-$0.61 MM1 disc in $\sim$0$\farcs$5-resolution SMA observations, compared to 77 and 166\,K for the $\sim$2$\farcs$4-resolution observations of \citet{cyganowski11a}.

On the whole, our results suggest that while the G19.01$-$0.03 MM1 disc could be stable, this would likely require the enclosed mass to be at the high end of our $40-70$\,M$_{\odot}$ range.
We thus speculate that, based on our data, it is more likely that G19.01--0.03 MM1 is unstable and may be undergoing fragmentation into as-yet undetected low mass stellar companions, as seen in G11.92--0.61 MM1 \citep{ilee18}.  
To test this hypothesis, we proposed high-resolution ($\sim$0\farcs09$\sim$370\,AU) ALMA observations.  These ongoing observations will have the spatial resolution and sensitivity to map the temperature structure of the candidate disc and search for evidence of fragmentation.

\subsection{The Nature of MM1}
\label{sec:nature_of_MM1}

The bolometric luminosity of G19.01--0.03 MM1 is $\sim$10$^{4}$\,L$_{\odot}$ \citep[based on fitting the MIR-mm wavelength SED;][]{cyganowski11a}.\footnote{The difference in assumed distance (4.0\,kpc in this work compared to 4.2\,kpc in \citealt{cyganowski11a}) corresponds to a decrease in luminosity of $\sim$9 per cent, within the uncertainty of the \citet{cyganowski11a} estimate (see their Figure 18b).} This is inconsistent with the enclosed mass of 40$-$70\,M$_{\odot}$ favoured by our kinematic modelling (\S \ref{sec:pv}) for the case of a single central object:  
if the stellar mass of $M_{\ast}=M_{\mathrm{enc}}-M_{d}\simeq33-65$\,M$_{\odot}$ corresponded to a single ZAMS star, the expected luminosity would be $>$10$^{5}$\,L$_{\odot}$ \citep[e.g.][Table 1]{davies11}.  The evidence for ongoing accretion in G19.01--0.03 MM1 (\S\ref{sec:intro}) means that the central source is unlikely to be in a ZAMS configuration, as discussed below.  Interestingly, comparison of observed MYSOs with candidate discs in the literature \citep[e.g.\ Table 7 of][]{johnston20} suggests that there is not a monotonic relationship between luminosity and enclosed stellar mass \citep[though luminosity estimates are affected by distance uncertainties, in particular for sources without maser parallax distances; see also][]{ilee16}.
Notably, a third of the sources tabulated by \citet{johnston20} are EGOs: G11.92$-$0.61 MM1, G23.01$-$0.41, G328.2551$-$0.5321 (EGO G328.25$-$0.53) and  IRAS 16547$-$4247 (EGO G343.12$-$0.06) \citep[][respectively]{ilee18,sanna19,csengeri18,zapata19}.  
While this is a small subsample, we note that even when considering only EGOs -- with similar MIR evidence for active outflows and so ongoing accretion -- there does not appear to be a monotonic relation between luminosity and enclosed mass, with G11.92$-$0.61 MM1 having the lowest luminosity \citep[$\sim$10$^4$\,L$_{\odot}$,][]{cyganowski11a,moscadelli16} and the highest stellar mass \citep[$>$30\,M$_{\odot}$, compared to $\le$20M$_{\odot}$ for the other sources:][]{ilee18,johnston20}.

It is now well understood that, in general, accretion affects the configuration (stellar radius and effective temperature) of accreting protostars \citep[e.g.][]{hosokawa09,hosokawa10,kuiper13,vorobyov17}.
\cite{hosokawa09}, \cite{hosokawa10}, and \cite{kuiper13} consider these effects for the high rates of mass accretion expected for high mass star formation.  These models show that as a protostar accretes solar masses of material, it is expected to undergo a swelling phase which causes the effective temperature to drop such that it is too low to produce an ionising flux sufficient for the creation of an \ion{H}{ii} region. Following the swelling phase, the protostar contracts and the effective temperature increases enough for ionisation. As recently shown by \cite{meyer19}, this is expected to happen in an episodic fashion in response to accretion outbursts, with MYSOs repeatedly experiencing episodes of bloating followed by "unswelling". 
As discussed by \citet{cyganowski11b}, the MIR-mm SED of G19.01$-$0.03 MM1 is well-fit by swollen, low-temperature models (T$<$5000\,K), though hotter (10,000\,K $<$ T $<$30,000\,K) models are also allowed (e.g.\ their Figure 6). 
As discussed in \S\ref{sec:sed}, the very small HC \ion{H}{ii} region implied by our SED modelling is consistent with gravitational "trapping" by an accretion flow; a swollen, non-ZAMS configuration would also contribute to the weakness of the observed centimetre-wavelength continuum emission (\S\ref{sec:continuum}, \S\ref{sec:sed}).

An enclosed stellar mass of $\sim$33$-$65\,M$_{\odot}$ corresponding to a single MYSO would place G19.01$-$0.03 MM1 among the most massive proto-O star candidates discovered to date.  Of the handful of sources with reported central stellar masses $>$30\,M$_{\odot}$ \citep[AFGL 2591-VLA3, G11.92$-$0.61 MM1, G17.64$+$0.16, G31.41+0.31:][respectively; see also \citealt{johnston20}]{jiminez-serra12,ilee18,maud19,beltran18}, only G17.64$+$0.16 (AFGL 2136) and AFGL 2591-VLA3 have central stellar masses $\ge$40\,M$_{\odot}$ and both have luminosities $\ge$10$^5$\,L$_{\odot}$. 
Theoretical models of accreting protostars also predict L$>$10$^5$\,L$_{\odot}$ for M$_*>$30\,M$_{\odot}$ \citep[e.g. Figure 12 of][]{kuiper13}.
This, together with the positional offset noted in \S\ref{sec:continuum}, motivates the consideration of another possibility: that the central stellar mass is instead distributed in a high-mass binary system.

The presence of a binary could potentially explain the offset (of $\sim$0\farcs16 $\sim$640\,AU; \S\ref{sec:continuum}) between the peak of the thermal dust emission and the VLA 5.01\,cm continuum emission.
One possible interpretation of the offset 5.01\,cm emission is an asymmetric \ion{H}{ii} region. 
In a single MYSO scenario, however, the centimetre emission from an \ion{H}{ii} region expanding into the cavities of the bipolar outflow \citep[e.g.][and references therein]{sartorio19,kuiper18} would be expected to be aligned with the outflow axis, which does not appear to be the case in G19.01 (Figure~\ref{fig:leaves}c).
The centimetre emission from an ionised jet would similarly be expected to be aligned with the bipolar outflow \citep[e.g.][]{guzman10}.
Interestingly, from the VLA 4.30 and 6.00\,cm images described in \S\ref{sec:sed}, there is a hint that the higher frequency (6.97\,GHz/4.30\,cm) emission may peak nearer the dust continuum than the lower frequency (5.00\,GHz/6.00\,cm) emission, as expected for collimated ionised jets \citep[e.g.][and observations of Cepheus A HW2 by \citealt{rodriguez94}]{reynolds86}. 
We caution, however, that from the present data it is unclear if this is a significant result: the difference in the 4.30 and 6.00\,cm positions is only $\sim$2.1$\times$ the uncertainty in the 6.00\,cm position, and the direction of the offset -- along an $\sim$E-W axis, with the lower frequency emission to the W -- is inconsistent with expectations for an ionised jet from a single MYSO, as outlined above. 
In the case of a binary system, an \ion{H}{ii} region and/or ionised jet could behave as expected with respect to an unresolved neighbouring source.

In sum, our results are 
consistent with the observation of a Keplerian disc $+$ infalling material, and of at least one MYSO given the high enclosed mass for all plausible inclination angles (\S\ref{sec:pv}).  
We note that adopting a higher (more nearly edge-on) inclination angle would reduce the inferred enclosed mass, reducing the mass-luminosity discrepancy (for the assumption of a single MYSO). 
Our ongoing 0\farcs09-resolution ($\sim$370\,AU) ALMA observations will help to illuminate the nature of MM1 by better resolving the disc and its kinematics, providing improved estimates of the disc inclination and enclosed mass.  These observations will also have sufficient linear resolution to detect wide binaries, such as the low-mass companion detected by \citet{ilee18} in G11.92$-$0.61 MM1 with a separation of $\sim$1920\,AU.  
Much closer binary companions are ubiquitous among O stars visible in the optical and NIR \citep[e.g.][who find 100 per cent of O dwarfs in their sample have a companion within 105 AU]{sana14}, and 
high-mass (proto)binaries with separations of 170 and 180\,AU have recently 
been observed in IRAS 17216-3801 and IRAS 07299-1651 \citep[][respectively]{kraus17,zhang19}.  
Searching for a comparably tight (proto)binary at the distance of G19.01$-$0.03 would require observations in the most extended ALMA configurations, highlighting the importance of long-baseline millimetre interferometric observations.

\section{Conclusions}
\label{sec:conclusions}

In this paper (Paper {\sc i}), we have presented a study of the nature and kinematics of the high mass (proto)star G19.01--0.03 MM1, using new subarcsecond-resolution ALMA 1.05\,mm and VLA 1.21\,cm and 5.01\,cm data.  Our main findings are as follows.

\begin{enumerate}

\item Compact molecular line emission detected with ALMA towards the MM1 millimetre continuum source exhibits a velocity gradient that is approximately perpendicular to the high-velocity bipolar molecular outflow driven by MM1. This velocity gradient is consistently traced by 20 lines of varying excitation energies of 8 molecular species, including the complex organic molecules (COMs) CH$_{3}$OH, CH$_{3}$OCH$_{3}$, g-CH$_{3}$CH$_{2}$OH, CH$_{3}$CH$_{2}$CN, and NH$_{2}$CHO.

\item Kinematic modelling shows that the observed velocities are well represented by a Keplerian disc model, including infalling material, with an enclosed mass of $40-70$\,M$_{\odot}$ within a 2000\,AU radius for an intermediate inclination angle of $i=40^{\circ}$ (estimated from the deconvolved sizes of the continuum and line emission).  This places G19.01--0.03 MM1 among the most massive proto-O star candidates with Keplerian discs to date.

\item A centimetre-wavelength counterpart to MM1, CM1, is detected for the first time in our VLA 1.21 and 5.01\,cm images.  Our modelling of the centimetre-millimetre wavelength SED confirms that thermal dust emission dominates at millimetre wavelengths, while a free-free component is required to explain the centimetre-wavelength emission.  The best-fit size of the ionised component (R$_e$=33 AU) is consistent with a small, gravitationally trapped hypercompact \ion{H}{ii} region. 

\item Our high-resolution observations show that the 6.7\,GHz Class II CH$_{3}$OH masers form a partial ellipse, consistent with an inclined ring with a fitted size of $\sim$600$\times$290\,AU, centred $\sim$0\farcs04 (160\,AU) north of the 1.05\,mm continuum peak.  The masers exhibit a velocity gradient consistent with that seen on larger scales in the thermal gas, suggestive of a rotational component to the maser motions.

\item We estimate a disc gas mass of $5.4-7.2$\,M$_{\odot}$ (for T$_{\rm dust}=$130-100\,K) from the observed 1.05\,mm flux density, assuming a simple model of isothermal dust emission.  This implies a central stellar mass (M$_{\rm enc}-$M$_{\rm disc}$) of $33-65$\,M$_{\odot}$.  Based on the disc-to-star mass ratio, our results indicate that the disc is likely to be unstable to fragmentation.

\item The bolometric luminosity of G19.01$-$0.03 ($\sim$10$^{4}$\,L$_{\odot}$) is lower than expected for a single accreting MYSO or ZAMS star with a stellar mass of $33-65$\,M$_{\odot}$.  This apparent discrepancy could be explained by multiplicity of the central source, with the mass distributed in an unresolved high-mass binary. 
The peak of the VLA 5.01\,cm emission is offset from the ALMA 1.05\,mm and VLA 1.2\,cm emission peaks by $0.16''\sim 640$\,AU, providing tentative evidence for a binary interpretation.

\end{enumerate}

In all, our results support the picture that G19.01--0.03 MM1 is a hot core source that harbours at least one MYSO, and potentially a high-mass binary system, 
which excites a small HC \ion{H}{ii} region and is fed by a Keplerian disc and ongoing infall. Higher angular and spectral resolution observations are required to further constrain the kinematic properties of the disc, search for disc fragmentation, and ascertain if MM1 hosts a high-mass binary system. Our ongoing ALMA observations, which can address many of these questions, will be presented in a future paper.


\section*{Acknowledgements}

We thank the referee, Rolf Kuiper, for a constructive report that helped clarify points in this manuscript. G.M.W. acknowledges support from the UK's Science and Technology Facilities Council (STFC) under ST$/$R000905$/$1 and ST$/$M001296$/$1, and wishes to thank Mark Thompson for useful discussion.  C.J.C. acknowledges support from the UK's STFC under ST$/$M001296$/$1 and
J.D.I. acknowledges support from the UK's STFC under ST$/$T000287$/$1.
J.M.D.K gratefully acknowledges funding from the Deutsche Forschungsgemeinschaft (DFG, German Research Foundation) through an Emmy Noether Research Group (grant number KR4801/1-1) and the DFG Sachbeihilfe (grant number KR4801/2-1), as well as from the European Research Council (ERC) under the European Union's Horizon 2020 research and innovation programme via the ERC Starting Grant MUSTANG (grant agreement number 714907).
This research has made use of: NASA's Astrophysics Data System Bibliographic Services, GILDAS (\href{https://www.iram.fr/IRAMFR/GILDAS}{https://www.iram.fr/IRAMFR/GILDAS}) and Python packages Astropy \citep{astropy}, astrodendro \citep{rosolowsky08dendro}, APLPY (\href{http://aplpy.github.com}{http://aplpy.github.com}), cmocean \citep{cmocean},  Matplotlib \citep{matplotlib}, NumPy \citep{numpy}, pandas \citep{pandas2010} and scitkit-image \citep{scikit-image}.
This paper makes use of the following ALMA data: ADS$/$JAO.ALMA$\#$2013.1.00812.S. ALMA is a partnership of ESO (representing its member states), NSF (USA) and NINS (Japan), together with NRC (Canada), NSC and ASIAA (Taiwan), and KASI (Republic of Korea), in cooperation with the Republic of Chile. The Joint ALMA Observatory is operated by ESO, AUI/NRAO and NAOJ. The National Radio Astronomy Observatory is a facility of the National Science Foundation operated under cooperative agreement by Associated Universities, Inc.


\section*{Data Availability}

The data underlying this article will be shared on reasonable request to the corresponding author.




\bibliographystyle{mnras}
\bibliography{bib} 





\bsp	
\label{lastpage}
\end{document}